\documentclass{elsarticle}

\usepackage{amsmath}
\usepackage{amssymb}
\usepackage{epsf}
\usepackage{graphicx}
\usepackage{dcolumn}
\usepackage{bm}

\newcommand{\nnabla}{\vec{ \nabla}}

\newcommand{\req}[1]{Eq.~(\ref{#1})}
\newcommand{\reqs}[1]{Eqs.~(\ref{#1})}
\newcommand{\rref}[1]{(\ref{#1})}

\newcommand{\x}{\mathbf{x}}

\newcommand{\n}{\mathbf{n}}
\renewcommand{\r}{\mathbf{r}}
\renewcommand{\v}{\mathbf{v}}

\newcommand{\beq}{\begin{equation}}
\newcommand{\eeq}{\end{equation}}
\newcommand{\be}{\begin{equation}}
\newcommand{\ee}{\end{equation}}
\newcommand{\beqa}{\begin{eqnarray}}
\newcommand{\eeqa}{\end{eqnarray}}
\newcommand{\bea}{\begin{align}}
\newcommand{\eea}{\end{align}}
\usepackage{amssymb}
\usepackage{array}
\usepackage{amsmath}
\usepackage{graphicx}\usepackage{graphics}
\usepackage{dcolumn}
\usepackage{bm}\usepackage{varioref}

\newcommand{\he}{\hat{\varepsilon}}

\renewcommand{\d}{\vec{d}}\newcommand{\f}{\vec{f}}
\newcommand{\R}{\boldsymbol{R}}

\newcommand{\sgn}{{\text{ sgn }}}

\newcommand{\Dt}{\left(\frac{d^2}{ dr^2}+\frac{d}{rdr}\right)}
\newcommand{\pt}{\partial_t}

\newcommand{\cH}{{\cal H}}
\usepackage{esint}

\begin{document}

\title{The internal structure of a vortex in a two-dimensional superfluid with long healing length and its implications}

\author[h]{Avraham Klein}
\address[h]{The Racah Institute of Physics, The Hebrew University of Jerusalem, 91904, Israel}
\author[c]{Igor L. Aleiner}
\address[c]{Physics Department, Columbia University, New York, NY 10027, USA}
\author[h,c]{Oded Agam}


\begin{abstract}
We analyze the motion of quantum vortices in a two-dimensional
spinless superfluid within Popov's hydrodynamic description.
In the long healing length limit (where a large number of particles are inside the vortex core) the superfluid dynamics is
determined by saddle points of Popov's action, which, in particular, allows for weak solutions of the Gross-Pitaevskii equation.
We solve the resulting equations of motion for a vortex moving with
respect to the superfluid and find the reconstruction of the
vortex core to be a non-analytic function of the force applied on the
vortex. This response produces an anomalously large dipole moment of
the vortex and, as a result, the spectrum associated with the vortex motion
exhibits narrow resonances lying {\em within} the phonon part of the spectrum,
contrary to traditional view.
\end{abstract}

\maketitle

\section{Introduction}

Hydrodynamics of superfluids, and in particular the dynamics of quantum vortices, have been studied extensively since the discovery of superfluidity in liquid Helium in the late 1930s \cite{Putterman74}. However, even today certain aspects of superfluidity are not fully understood. For instance, the problems of the vortex mass and of the force acting on it are still controversial (see Ref.~\cite{Sonin2013} and references therein), and there is no full theoretical understanding of the excitation spectrum of superfluids (for recent works on this subject see, e.g., \cite{Henkel2010,Arrigoni2013}).

Interest in the theory of superfluidity has revived considerably in recent years, ever since atom cooling techniques opened up new experimental avenues for realization of superfluids using Bose-Einstein condensates (BEC) of cold atoms \cite{Anderson95,Bradley95,Davis95}. These systems
 provide precise tomography of the condensate structure, and are used nowadays to gather additional information on problems which were studied, in the past, using superfluid Helium (for a review see \cite{vinen02} and references therein). Examples of these are dynamics of vortices \cite{Matthews99,Madison00,Weiler08,Freilich10, Neely10}, Kelvin waves \cite{Smith04}, vortex lattices \cite{AboShaeer01,Schweikhard04}, Tkachenko waves \cite{Coddington03}, vortex tangles \cite{Henn09}, and quantum turbulence \cite{Neely13}.

However, whereas liquid Helium represents a superfluid in the strong coupling regime, BECs are usually described by a weakly coupled theory, as the strength of interatomic interactions can be controlled and tuned to be very small. One difference between these two types of systems manifests itself in the size of the vortex core. In the strong coupling regime the vortex core is, effectively, of the interatomic distance, while in the weak coupling limit it is much larger. The core size of the vortex has, in turn, important implications for the vortex dynamics. For vanishingly small vortex cores, the Kelvin circulation theorem \cite{Lamb} implies that a vortex can only move together with its surrounding fluid, i.e. one cannot apply a force on a vortex.
In the other extreme, i.e. when the vortex core is very large compared to the interatomic distance, forces acting on vortices are non-perturbative. They may deform the vortex core and play an important role in the dynamics of the vortex.  In other words the hydrodynamical description of a superfluid with vortices should include not only the positions and vorticity charges of the vortices, but also (at the very least) their dipole moments.

Most studies of quantum vortices have employed the mean field description given by the Gross-Pitaevskii equation (GPE), which provides a good approximation in the weak coupling regime. However, the question concerning the internal structure of the vortex and, in particular, its description by weak solutions of the GPE has largely been overlooked. (By weak solutions we refer to the case were the GPE is satisfied everywhere in space except for a set of zero measure).

In this paper we show that these weak solutions necessarily appear when a force is applied to a vortex. Moreover, they manifest themselves in an anomalously large dipole moment. We construct the hydrodynamical description of a spinless superfluid in two dimensions, taking into account the dipole moment of the vortices. This description applies over distances larger than the vortex core and when vortices are far apart. A prominent prediction of this effective theory is the existence of low-energy excitation-levels of vortex core. We show that these energy level are long-lived, and characterize their influence on the elastic scattering of phonons from vortices.

The rest of the paper is organized as follows. In the following section we present a qualitative discussion of our result. In Sec. \ref{sec:popovs-formalism}, we review Popov's field theory \cite{PopovBook} of superfluids where vortices are introduced as constraints. Then, in Sec. \ref{sec:solution}, we present our solution of the Popov equations for the case of a pinned vortex in a uniform flow. We first show the numerical solution of the problem and then provide analytical understanding by solving the Bogolyubov equations and matching the solution to the weak solution of the GPE in close vicinity to the vortex core.  Next, in Secs. \ref{sec:eft} and \ref{sec:cm}, we construct the effective field theory which takes into account the dipole-moment of the vortex, derive the corresponding equations of motion, and solve them in order to describe the classical oscillatory motion of a vortex and its coupling to phonons. In Sec. \ref{sec:qm} we quantize the vortex motion. Finally, in Sec. \ref{sec:conclusions}, we conclude and discuss some future directions of the study of the non-analytic core deformation of vortices.

\section{Qualitative discussion and summary of the results.}
\label{sec:qual-disc-summ}

The hydrodynamic equations describing stationary solutions of a superfluid far from vortex cores are
the conservation of mass,
\begin{subequations}\label{hydrodynamic}
\be
\vec{\nabla} \cdot (n \vec{v}\,) =0,
\label{hydrodynamica}
\ee
where $n$ and $\vec{v}$ are the superfluid density and velocity respectively, and the requirement of potential flow outside the vortex,
\be
\vec{\nabla} \times \vec{v}  =0.
\label{hydrodynamicb}
\ee
Finally, for potential flow and zero-entropy evolution, the Bernoulli
equation (in the local approximation for the equation of state) reads,
\be
\vec{\nabla} \left( \frac{ \vec{v}^{\, 2}}{2} + n \right) =0. \label{Bernoulli}
\ee
\end{subequations}
Here, and from this point onwards, we work in dimensionless units such that density is measured in units of the equilibrium
density, $n_0$, the length is measured in units of the healing length
$\xi= 1/\sqrt{\lambda n_0}$ where $\lambda$ is the dimensionless coupling constant, and the velocity,
$\vec{v}$, is measured in units of $\hbar/(m\xi)$ where $m$ is the particle's
mass. Accordingly, the frequency is measured in units of healing frequency
$\hbar/(m\xi^2)$, and the energy is measured in units of
$\hbar^2/(m\xi^2)$. The speed of sound in our units is one. The weak coupling limit, on which we focus, is realized
when $1/\lambda = n_0 \xi^2 \gg 1$, namely when the number of particles within a square of size of the healing length squared is much larger than one.

For the sake of convenience we shall also represent vectors in the
complex plane, e.g. $v=v_x+i v_y$.
In this representation the solution of a single vortex, with unit
vorticity, located at the origin is given by $v= i/z^*$,
where $z=x+iy$ is the complex coordinate and $(\cdot)^*$ denotes complex
conjugation. From \req{Bernoulli} it thus follows that the density is
\be
n = 1- 1/(2 |z|^2)
\label{density1}
\ee
sufficiently far from the vortex core, i.e. for $|z| \gg 1$ (at short
distances the gradient terms in the equation of state become important
and will be discussed later).

Consider, now, a single vortex moving with constant
velocity $-\vec{v}_c$, and let us choose a moving coordinate frame attached to the vortex
core (it corresponds to a fixed vortex and a superfluid flow
$\vec{v}_c$  far from the vortex). If one could ignore \req{Bernoulli}
and require $n=1$ (i.e. an incompressible liquid model),
the solution of the first two equations of Eqs. (\ref{hydrodynamica}--\ref{hydrodynamicb})
would be a superposition of the two
contributions to the superfluid velocity: $v\equiv v_x+iv_y=
i/z^*+v_c$, or in other words, the vortex moves with the flow.
However, the nonlinear coupling between the velocity and density \rref{Bernoulli}, modifies the solution such that, to linear order in $v_c$, one finds:
\be
v \simeq  v_c+\frac{i}{z^*}+\frac{id^*(|z|)}{(z^*)^2} +
{\cal O}\left(\frac{v_c}{|z|^2}\right), \quad |z| \gg 1,
\label{dipole1}
\ee
where $d=d_x+id_y$ is the dipole moment describing the deformation of
the vortex-density shape under the effect of the external flow,
\be
d=-iv_c \ln|z| + d(1)
\label{dipole2}
\ee
Equations (\ref{dipole1} -- \ref{dipole2}) show that the virtual position of the vortex core,
when viewed from far distances, is shifted with respect to its actual
position by $d$. This shift is proportional to the velocity of the vortex and
grows logarithmically with the distance. The same shift  is also manifested in the superfluid density around the vortex.

One of the main conclusions of the present paper concerns the dependence of
the shift $d(1)$ on the superfluid velocity $v_c$. Naively, one could
estimate  $d(1)\simeq -i\beta v_c$, where $\beta$ is a constant of the
order of unity, so it does not have any visible effect on the
background of the logarithmic function. However, contrary to this naive view, we show that
the dipole moment has a non-trivial and non-analytic dependence on the
velocity,
\be
d(1)= -\frac{iv_c}{\alpha^2}\ln\frac{1}{|v_c|},
\label{dipole3}
\ee
where $\alpha=0.8204\dots$ is a constant that describes the density profile near a stationary
vortex ($v_c=0$),
\be
n=\alpha^2|z|^2 + \dots,\quad |z|\ll 1.
\label{n-small-z}
\ee

\begin{figure}[t]

\includegraphics[width=0.72\columnwidth]{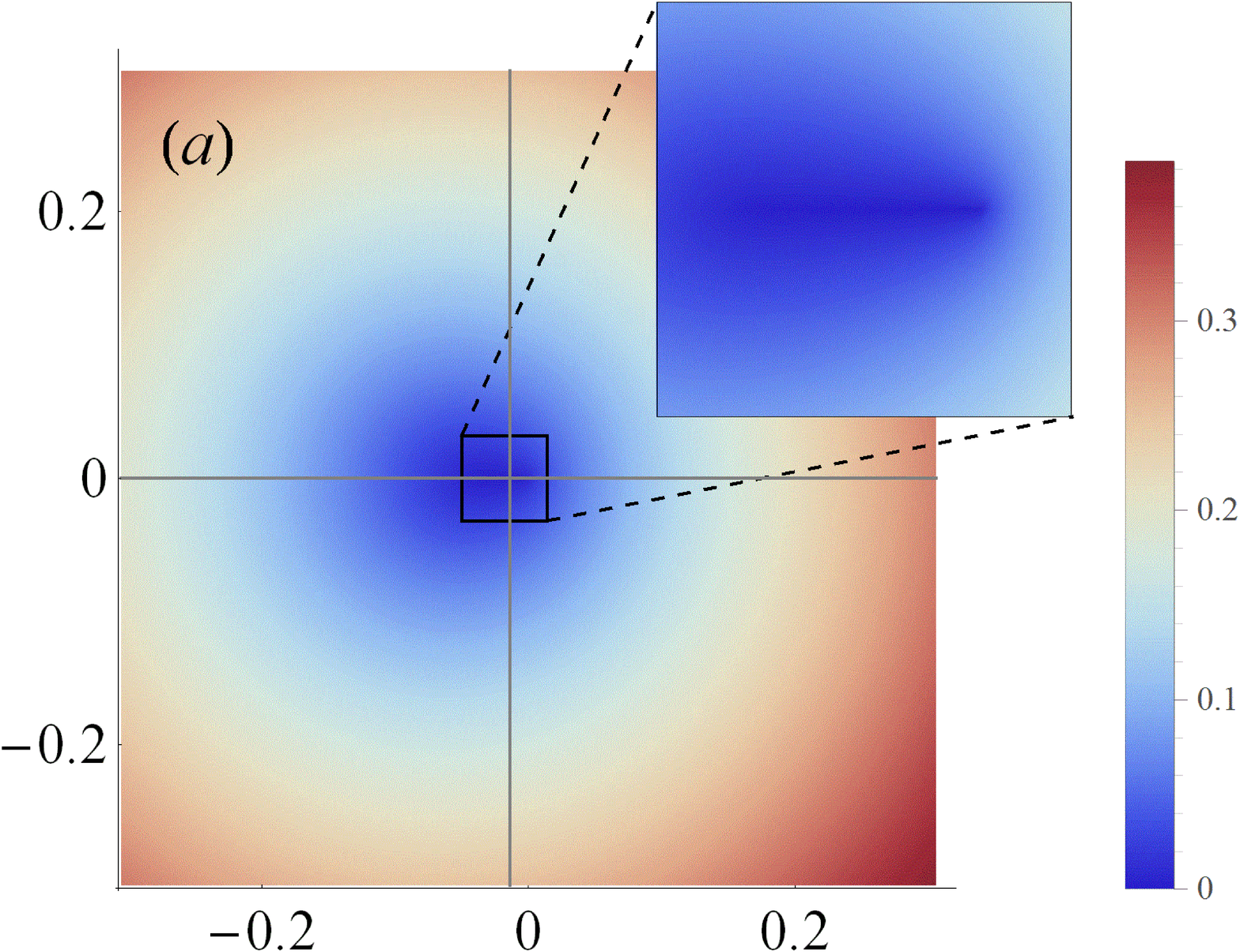} \\[3mm]
\includegraphics[width=0.47\columnwidth]{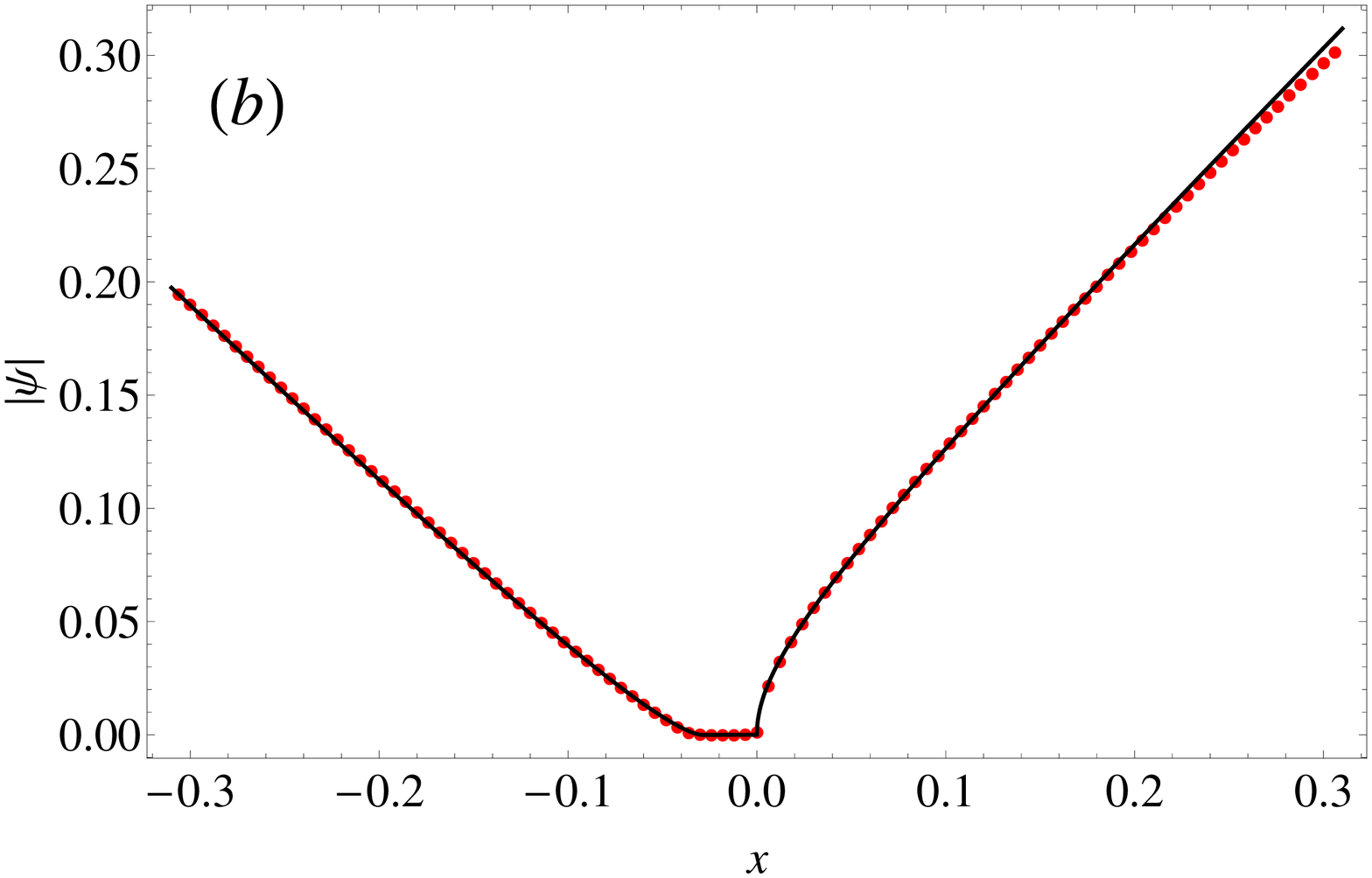}  
\includegraphics[width=0.47\columnwidth]{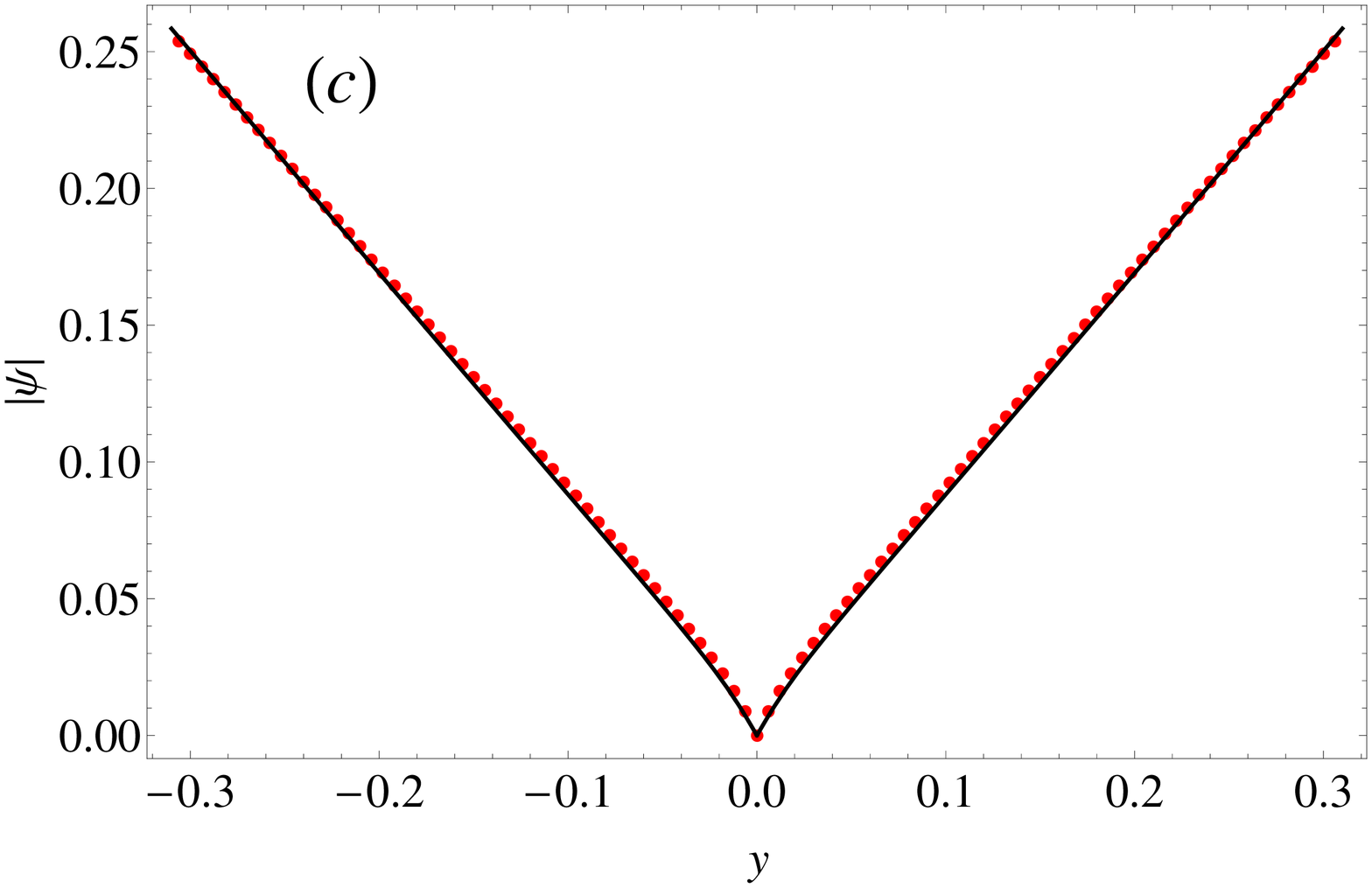}
\caption{The square root of the density, $|\psi|$, of a superfluid
  near the core of a moving vortex
(either vortex moves in positive $y$-direction or the supercurrent
is along negative $y$-direction for a vortex in rest)
: (a) Two dimensional density plot.
(b) A section of $|\psi|$ along the horizontal gray line in upper panel. (c) A section of $|\psi|$ along the vertical gray line of the upper panel which bisect the cut at its midpoint. The red dots are the result of our numerical solution while the solid line is obtained from formula (\ref{cut}).}
\label{fig1}
\end{figure}

The origin of the logarithmic dependence of dipole moment \rref{dipole3} is in a so
called {\em weak} solution of the stationary Gross-Pitaevskii equation which is
obtained from the variation of the functional:
\be
E\left[\psi,\psi^*\right]=
\int d^2 r  \left[\frac{|\nabla \psi|^2}{2} + \frac{(|\psi|^2-1)^2}{2}\right]. \label{action}
\ee
Numerical solutions of the corresponding stationary equations
(obtained with the help of Popov's formalism which will be described
in Sec.~\ref{sec:Popov})
demonstrate that the logarithmic shift of the virtual position of the vortex
indeed emerges from the singular structure of the vortex core.
 Figures~\ref{fig1} shows the plots for density profile. 
The solid line in Figs.~\ref{fig1} b,c)
 is the following analytic solution for distances
much smaller than the healing length
\begin{subequations} \label{cut}
\be
\psi = \frac{i v_c}{2 \alpha} \left[ Z-\frac{1}{Z^*} +\ln (Z Z^*) \right],
\ee
 where
\be
Z= \frac{\alpha^2 z}{i v_c} +\sqrt{\left(\frac{\alpha^2 z}{iv_c}\right)^2 -1}, ~~~~\mbox{and}~~~ |z| \ll 1.
\ee
\end{subequations}
(The function $\sqrt{\zeta^2-1}$ is defined to be analytic in the entire complex plane except for a cut along the real axis
between the points $\zeta=-1$ and $\zeta=1$.) This solution is obtained by approximating the Gross-Pitaevskii equation with the
Laplace equation, see Sec.~\ref{sec:solution}.
 From here it follows that the solution of a moving vortex
 exhibits a non-analytic behavior in a form of a cut of
length proportional to the velocity $v_c$. One branch point of the cut
is located at the point in which the vortex is pinned,
and the cut extends in a direction perpendicular to the velocity, $v_c$, i.e. in the same direction as the Magnus force acting on the vortex.

This nonanalytic solution of the Gross-Pitaevskii equation should be
understood as a {\em weak solution}. Namely, in similar manner to shock waves in
hydrodynamics, it represents a saddle point of the energy
(\ref{action}) but it is not an analytic solution of the saddle point's
equations. The weak solution is valid only outside some limited region
of measure zero (the vicinity of the cut in our case). Within that region the
description of the superfluid by the functional \rref{action} does not
hold, and additional physics should be taken into
account by proper regularization. Nevertheless,
the weak solution implies that, contrary to the
traditional view, a vortex in the weak coupling limit possesses an
anomalous contribution to a dipole-like degree of freedom which is associated with the
length and the direction of the cut. The physical results
which we will obtain from this property do not depend on the regularization scheme at all.

The appearance of the anomalous dipole moment \rref{dipole3} has
a profound effect on the low-energy dynamics of the vortices.
We shall enumerate these effects here and
relegate their detailed discussion to the corresponding
sections:
\begin{enumerate}
\item
The dipole moment leads to similar 
non-analytic dependence of the {\it added mass}
(see, e.g., Ref.~\cite{Hydrodynamics}) of the vortex, $M_v\simeq \hfill m (n_0\xi^2)\ln (1/|v|)$ where $|v|$ is the velocity of the vortex with respect to the
superfluid, see Sec.~\ref{sec:cl-em} for more details.

\item
Under the effect of the Magnus force, the vortex exhibits a periodic
circular motion with frequency $\Omega\propto \frac{\hbar
  n_0}{M_v}\simeq \frac{\hbar}{m\xi^2\ln (1\|v|)}\ll
\frac{\hbar}{m\xi^2}$. Thus, vortex oscillations with
small amplitude have frequencies within the phonon spectrum, contrary to the widely accepted view in the literature, see Sec.~\ref{sec:cm2}.

\item The circulating vortex  emits sound waves (phonons) but this emission is
  suppressed by an additional factor of $1/[\ln (1/|v|)]^3$, see
  Sec.~\ref{sec:cm4}. Therefore, the excited states of the vortices
  are well defined resonances.

\item Quantization of the periodic motion of the vortex, leads to quantization of the radius of the circular motion which determines, in turn, the discrete energies of those resonances. These resonances lie within the phonon spectrum, provided that the superfluid has a long healing length, $n_0\xi^2 \gg 1$, see Sec.~\ref{sec:qm}.
\end{enumerate}

\section{Popov's formalism.}
\label{sec:popovs-formalism}

\label{sec:Popov}

In this section we briefly review the derivation of Popov's representation of the dynamics of a two-dimensional compressible
superfluid \cite{PopovBook}.

To shorten the notation, we represent vectors in the two dimensional space
using arrow, e.g. $\vec{r}=(x,y)$, or $\vec{v}=(v_x,v_y)$. We also use 3-vectors notation in which the
first component is time-like and the two other are space-like.
These will be represented by boldface letters, e.g. ${\bf x}=(t,x,y)$,
and $\boldsymbol{\partial} = (\partial_t, \partial_x,\partial_y)$.

All zero-temperature quantum  properties of the superfluid can be
obtained by a suitable differentiation of the ``partition function''
\be
{\cal Z}=\int {\cal D}\psi^* {\cal D}\psi\exp\left(\frac{i{\cal S}}{\lambda}\right),
\label{Partition}
\ee
with the action
\be
{\cal S}=\int d^3 x i\psi^*\pt\psi-\int dt E\left[\psi,\psi^*\right],
\label{actionA}
\ee
and the energy functional $E\left[\psi,\psi^*\right]$ given by \req{action}. The interaction constant
$\lambda \ll 1$, therefore, plays the role of the Planck constant which
allows for the semiclassical treatment of the problem in the weak coupling regime.

\subsection{Derivation of Popov's action.}

The derivation of Popov's theory of superfluids is comprised of two mains steps.
The first is the Madelung transformation:
\be
\psi = \sqrt{n} \exp(i\theta),
\label{psi}
\ee
where $n$ is the density of the superfluid and $\theta$ is its phase,
which transforms the action (\ref{actionA}) to
\be
{\cal S}= - \int d^3 x  \left[ v_0 n + \frac{n
    |\vec{v}|^2}{2} + \frac{ |\vec{\nabla} n|^2}{8n} +
  \frac{(n-1)^2}{2}\right].
 \label{action1}
\ee
where ${\bf v}=\left(v_0,v_x,v_y\right)$ are given by
\be
{\bf v} = \boldsymbol{\partial} \theta.
\label{constraint1}
\ee

The well-known disadvantage of this representation of \rref{psi} is that
$\theta$ is a multi-valued function of the coordinate, and therefore taking into account
configurations with moving vortices  is somewhat
inconvenient. The second step of the derivation introduces new single-valued fields and vortices as a way to resolve \req{constraint1} as well as
the multi-valuedness of the field $\theta$.

We consider, instead of $\theta$, a three-component field of the
velocity, $ \v(\x)$. According to \req{constraint1} these components
are not independent,
\be
\epsilon_{ijk} \partial_j v_k =2\pi J^V_{i}\label{constraint2}
\ee
where $\epsilon_{ijk}$ is the unit antisymmetric tensor. In the
 right hand side of \req{constraint2} are the components of the 3-vector representing the vortex current. These take into account that $\theta$ is not a
single-valued function and can be changed by the multiples of $2\pi$
while traveling along any contour encompassing a vortex line $R_l(t)$:
\be
J^V_0 =\sum_{l}\sigma_l\delta[\vec{r}- \vec{R}_l(t)]; \quad \vec{J}^V =\sum_{l}\sigma_l\pt{\vec{R}}_l\delta[\vec{r}- \vec{R}_l(t)].
\label{vortexcurrent}
\ee
Here, $\vec{R}_l(t)$ and $\sigma_l$ denote the position and the vortex charge of the $l$-th vortex. The vortex current is analogous to the particle  current in the usual
electrodynamics and, by construction, satisfies the continuity
equation
\be
 \boldsymbol{\partial}\cdot\boldsymbol{J}^V=0.
\label{vortex-continuity}
\ee

Introducing gauge fields $\boldsymbol{A}=(a_0,-a_x,-a_y)$ to resolve
constraint \rref{constraint2}, we rewrite \req{Partition} as
\be
{\cal Z}=
\left[\prod_l \int  {\cal D}\R_l(t)\right]
\int {\cal D}n {\cal D}\v {\cal D}\boldsymbol{A}
\exp\left[\frac{i{\cal S}}{\lambda}+\frac{i}{\lambda }\int d^3 x
{A}_i\left(\epsilon_{ijk} \partial_j v_k - 2\pi J_{i}^V\right)
\right],
\label{partition2}
\ee
where action ${\cal S}$  is given by \req{action1}.

The integration over all three components of the velocity $\v$ in
\req{partition2} can be performed exactly, leading to Popov's
non-linear electrodynamics:
\be
{\cal Z}=
\left[\prod_l \int  {\cal D}\R_l(t)\right]
\int {\cal D}\boldsymbol{A}
\exp\left[\frac{i{\cal S}_P}{\lambda}
\right],
\label{partition3}
\ee
where the vortex current is defined in \req{vortexcurrent} and Popov's
action is given by
\be
{\cal S}_P=
\int d^3 x  \left[ \frac{\vec{E}^2}{2B} - \left(\frac{ |\vec{\nabla}
    B|^2}{8B} + \frac{(B-1)^2}{2}\right)- 2\pi \boldsymbol{J}^V \cdot \boldsymbol{A}\right]. \label{action2}
\ee
Popov's field  $\vec{E}=(E_x,E_y)$ and $B$ are analogous to electric
and magnetic fields in two-dimensional electrodynamics
\be
B=\nnabla\times \vec{a};\quad \vec{E}=-\pt \vec{a}-\vec{\nabla}a_0.
\label{fields}
\ee
 The last term of the action (\ref{action2}) has the form of the interaction of the particles (vortices) with the fields. The gauge invariance of the theory \rref{partition3} is guarded by
the vortex current conservation \rref{vortex-continuity}.

The physical superfluid current $\vec{j}$ and density $n$ can
expressed in terms of Popov's fields as
\be
\left(n,j_x,j_y\right)=\left(B,E_y,-E_x\right).
\label{Popov-physical}
\ee

\subsection{Popov's equations}

Let us now write down the equations of motion which follow from Popov's formalism.
The first dynamical equation, analogous to  Faraday's law, is obtained directly
from the relation (\ref{fields}) of the fields, $\vec{E}$ and $B$, to the gauge fields, $\bf{A}$:
\begin{subequations}\label{popov11}
\be
\partial_t B=-\nnabla\times \vec{E}.
\label{popov11a}
\ee
From comparison of \req{popov11a} with \req{Popov-physical} it follows that this equation is just the requirement of conservation of physical mass charge $\pt
n+\nnabla\cdot\vec{j}=0$.

The other Maxwell equations are obtained by varying of the action
\rref{action2} with respect to the vector-potential $\boldsymbol{A}$.
Variation with respect to $a_0$ gives an analogue of Gauss' law:
\be
\nnabla \cdot \left(\frac{\vec{E}}{B}\right) =2\pi J^V_0(\r),
\label{popov11b}
\ee
where vortex density and currents are defined in \req{vortexcurrent}.
This equation replaces the quantization condition of the velocity circulation.
The third equation, analogous to Amper's law, is obtained by variation of the action with respect to $\vec{a}$:
\be
\he\nnabla \left[
 \frac{\vec{E}^2}{2 B^2}+B-\frac{1}{2\sqrt{B}}\nnabla^2\sqrt{B}
\right] =
\vec{J}^{\,V}(\r,t)+\partial_t \left(\frac{\vec{E}}{B}\right),
\label{popov11c}
\ee
where $\he$ is an antisymmetric tensor of the second rank acting on
the spatial coordinates.
\end{subequations}

A further advantage of the action \rref{action2} is that it allows one to
immediately find the force acting on a vortex. Varying the action with
respect to the vortex position, we find
\[
\sum_l\int dt \vec{F}_l\delta \vec{R}_l=-\delta \left(
 2\pi\int d^3\x \boldsymbol{J}^V \cdot \boldsymbol{A}\right).
\]
Direct calculation utilizing \reqs{fields} gives
\be
\vec{F}_l=2\pi\sigma_l \vec{f}_l;
\quad \f_l\equiv
\vec{E}(\vec{R}_l,t)
+  B(\vec{R}_l,t) \he \pt \vec{R}_l
\label{force1}
\ee
which is equivalent to the usual expression in linear electrodynamics.

Equation \rref{force1} is just an expression for the Magnus
force provided that $\vec{E}$ and $B$ are understood as external fields which
do not include the fields created by the vortex itself. In
standard electrodynamics such exclusion of the self-interaction is
easily performed because the equations of motion for the fields are linear
and the superposition principle holds. Here, Eqs.~(\ref{popov11a}--\ref{popov11b}) are non-linear and exclusion of the self-interaction is not a trivial task. We postpone our consideration of this question until we have had a chance to study in detail the fine structure of
the core of a moving vortex (in Sec.~\ref{sec:solution}),
and to obtain the equation of motion for the vortex (in Sec.~\ref{sec:eft}).

\subsection{Galilean invariance}

Popov's action and equations of motion preserve the Galilean invariance
of the original model. One can show that the transformations
\be
\begin{split}
&
a_0(t,\vec{r}) \to a_0(t,\vec{r}-\vec{v}t)+\vec{v}\cdot \vec{a}(t,\vec{r}-\vec{v}t); \quad \vec{a}(t,\vec{r}) \to \vec{a}(t,\vec{r}-\vec{v}t);
\\
& \vec{B}(t,\vec{r}) \to \vec{B}(t,\vec{r}-\vec{v}t); \quad
 \vec{E}(t,\vec{r}) \to \vec{E}(t,\vec{r}-\vec{v}t)
-B(t,\vec{r}-\vec{v}t)\he \vec{v};
\\
& \vec{R}_i(t)\to  \vec{R}_i(t) + \vec{v};
\\
&
J^V_0(t,\vec{r})\to J^V_0(t,\vec{r}-\vec{v}t);
\quad \vec{J}^V(t,\vec{r})\to \vec{J}^V(t,\vec{r}-\vec{v}t)
+\vec{v} J^V_0(t,\vec{r}-\vec{v}t)
;
\end{split}
\label{Galileantransform1}
\ee
(with $\vec{v}$ being any constant velocity) leave equations of motion
\rref{popov11} and the force acting on a vortex \rref{force1}
intact. Under the same transformation, Popov's action acquires an extra term
\be
\delta {\cal S}_P=\int d^3 x \left[
\frac{\vec{v}^{\,2}B}{2}-\vec{E}\he\vec{v}\right],
\label{ActionGalilean}
\ee
which is just the change of the total kinetic energy in the
moving frame. As each of the terms is a total derivative, they
do not influence the dynamics of the system.

\subsection{Relation to the Gross-Pitaevskii equation.}

Everywhere in space, where there are no vortices and $B>0$, it is possible to express the solution
of \reqs{popov11} in terms of the Gross-Pitaevskii equation
\be
i\pt\Psi=-\frac{1}{2}\nnabla^2\Psi+\left(|\Psi|^2-1\right) \Psi.
\label{GP}
\ee

Indeed, substituting
\be
B=|\Psi|^2;\quad \vec{E}= \frac{i}{2}\he\left[\Psi^*\nnabla \Psi
  -\Psi\nnabla \Psi^* \right],
\label{BEGP}
\ee
into \reqs{popov11}, we find those equation are satisfied wherever the
vortex density and current equal zero. The advantage of Popov's
variables is that they allow us to include configurations where equation
\rref{GP} is not satisfied on a set of measure zero, or in other words weak solutions of the GPE.
An example of this situation is a pinned vortex in a superfluid flow, which we describe in the next section.

\section{Solution of the Popov equations for a pinned vortex in
  a superfluid flow}
\label{sec:solution}

In this section we discuss the solution of Popov's equations for a pinned vortex in an otherwise constant superfluid flow. We first present our numerical procedure for the solution of this problem, and show that  it is, in fact, a {\it weak solution} which exhibits a cut singularity within the vortex core.  Next we provide an analytic interpretation of the result. To this end we first solve the Bogolyubov equations which provide the perturbative solution of the problem far from the cut-singularity. Then we solve the Gross-Pitaevskii equation deep within the vortex core by matching the solution to the perturbative result.

As we are interested in a stationary solution of Eqs.~(\ref{popov11}),
it may be described by two real fields, $\rho=\sqrt{B}$ and $a_0$
(the electric field is given by $\vec{E}= -\vec{\nabla} a_0$), thus Popov's equations reduce to the following pair of equations:
\begin{subequations}\label{EQ:numerics}
\beqa
\label{eq:sGPE-1}
-  \vec{\nabla} \cdot \frac{\vec{\nabla} a_0}{\rho^2} &=& 2\pi\sigma \delta(\vec{r} - \vec{R}_0)\\
   \hat{H} \rho &=& 0\label{eq:sGPE-2}
\eeqa
where ${\hat H}$ is a nonlinear operator which is defined as
\begin{equation}
  \label{eq:GPE-popov-hamilt}
   \hat{H} \rho = \left(\frac{1}{2}\vec{\nabla}^2 - \frac{1}{2}\left|\frac{\vec{\nabla} a_0}{\rho^2}\right|^2 + 1 - \rho^2\right)\rho.
\end{equation}
\end{subequations}
The boundary conditions on $\vec{E}$ are that far from the vortex the superfluid current is constant, namely $\vec{E}(r\to\infty)=-\he \v_c$.

\begin{figure}
\vspace*{-3cm}
\begin{center}
\includegraphics[width=0.8\columnwidth]{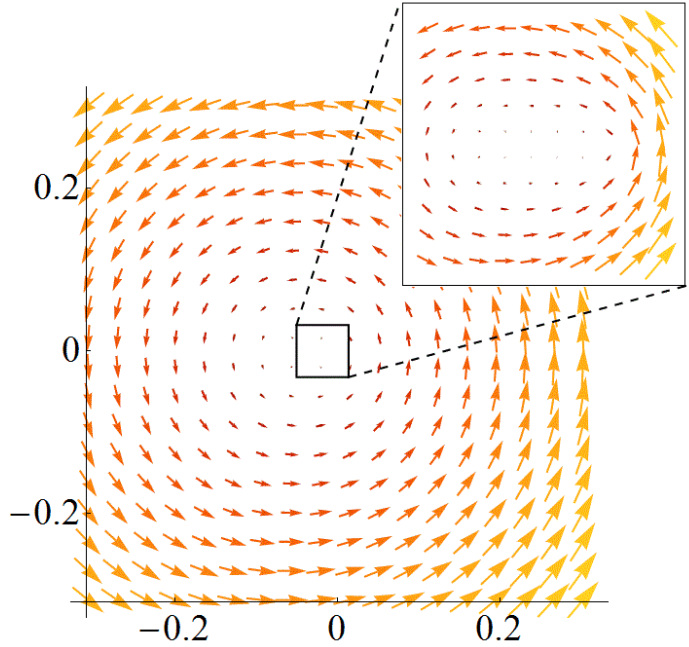}
\caption{The superfluid current in the vortex core of a pinned vortex in a uniform flow, $\vec{j}=\hat{\epsilon} \vec{E}$. The inset shows a magnified view of the region near the cut similar to Fig.~\ref{fig1}.}
\end{center}
\label{fig2}
\end{figure}

\subsection{Numerical solution} \label{sec:numer-analys-popovs}
The numerical results for the solution of the above equations for the square root of the density and the superfluid current in the vicinity of the vortex core are depicted in Fig.~\ref{fig1}, and Fig.~\ref{fig2}, respectively. Here we sketch the numerical procedure used in order to obtain these results. Additional details can be found in Appendix A.

\begin{figure}
\vspace*{-3cm}
\begin{center}
\includegraphics[width=0.8\columnwidth]{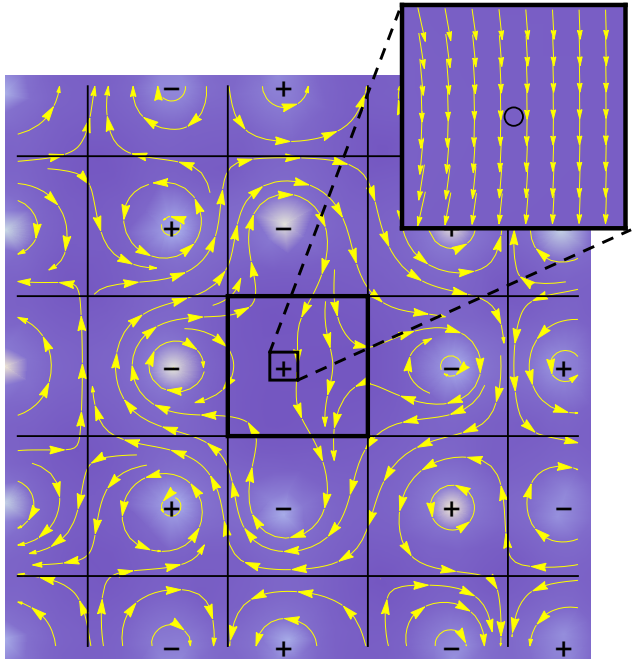}
\caption{The lattice of image vortices generated due to Dirichlet boundary conditions, and the effective current (total current minus the current created by the central the vortex) produced when the vortex is shifted from the central point of the domain. The inset shows a magnified view of the region of the vortex. Here, the circle represents the size of a vortex, i.e. a circular region with radius equals to the healing length. This size is much smaller that the scale over which the flow changes.}
\label{fig3}
\end{center}
\end{figure}

Working with two real fields, $a_0$ and $\rho$, simplifies the
numerical task. For the numerical procedure it is efficient to apply
the external superfluid flow by placing the vortex in a very large square box and slightly
shifting its position from the center of the box. The boundary conditions on the box are equivalent to the creation of ``image'' vortices and anti-vortices. The shift of the vortex from the center results in a  superfluid flow at the position of the vortex, as illustrated in Fig.~\ref{fig3}.
 To see how this works, consider the
boundary conditions of a superfluid in a box. The component of the
superfluid current normal to the boundaries must be zero, and
therefore the electric field must be normal to the boundaries. Thus we
have the electrostatic problem of a grounded metal box, with Dirichlet
boundary conditions:
\begin{subequations}
\begin{equation}
  \label{eq:a0-boundary}
  a_0(x,\pm L/2)=a_0(\pm L/2,y)=0.
\end{equation}
where $L$ is the size of the box. In analogy with the electrostatic
problem, we can remove the boundaries by creating a lattice of image
``charges'', i.e. vortices with alternating vorticities. The superfluid density is
the same for a vortex and antivortex, thus we need periodic boundary
conditions for $\rho$. Since $\rho \geq 0$, we chose to use the
equivalent and slightly more convenient Neumann boundary conditions
\be\label{eq:rho-boundary}
\hat{n}\cdot \vec{\nabla} \rho\Bigg|_{\vec{r} \in\ {\rm boundary}}=0,
\ee
where $\hat{n}$ is the normal to the boundary. Shifting the vortex core from the center of the box, say by choosing $\vec{R}_0 = (-\Delta x, 0)$, shifts and deforms the entire lattice and imposes a flow at $\vec{R}_0$, see Fig.~\ref{fig3}. This flow can be calculated to excellent numerical accuracy by solving the analogous electrostatic problem.
\end{subequations}

Equations such as \reqs{EQ:numerics} can be solved via relaxation to equilibrium of the fictitious time-dependent problem:
\begin{equation}
  \label{eq:GPE-time}
  \hat{H} \rho = \frac{\partial \rho}{\partial t}
\end{equation}
The dynamical system of Eqs.~\eqref{eq:sGPE-1}--\eqref{eq:GPE-time} has a convenient fluid-dynamical interpretation. Essentially, the equations describe diffusive dynamics of an interacting fluid. The fluid is vorticity-free except at a single point $\vec{R}_0$.

We solve the equations by discretizing the space and the time, iterating \req{eq:GPE-time} and enforcing \req{eq:sGPE-1} explicitly at the end of each time step, i.e. solving the resulting linear equation for $a_0$ for the temporal value of $\rho$. This procedure is similar that used in hydrodynamic problems (see e.g. Ref.~\cite{Ames1992} for an overview of this topic).

Special consideration must be given to the discrete form of
\req{eq:sGPE-1}, which describes a conservation law. Its discrete form should enforce the discrete
version of the law, or in other words, the discrete form of the
current conservation $\vec{\nabla} \times E =0$
(which follows from \req{popov11a} for stationary case).
A convenient way to achieve this, when the dynamics are given by an action, is to to derive the equation by variation of the discretized action \cite{Adler1984}. A full description of the algorithm may be found in~\ref{appendixA}.

Finally, to deal with the singular behavior that emerges in the vicinity of the vortex location, $\vec{R}_0$, one  needs fine discretization near the vortex, usually going to lattice constants of order $\Delta \sim 0.01 - 0.001$. On  the other hand one should also maintain large system sizes, usually $L \sim 50 - 100$, which implies an excessively large number of lattice points. To resolve this problem we used multiple-resolution grids, often using 3 grids so that near the core singularity we had very high resolution, sometimes as high as $\Delta = 0.0005$. We interpolated the grids to each other using bicubic splines.

There is no general method of insuring stability and correctness for nonlinear problems with singularities. Instead, we monitored the progress of the iteration and checked that the solution converged properly. In order to increase accuracy, we made sure that a high-resolution grid always surrounded $R_v$ in order to reduce the so called ``infection region'' of the singularity.

\subsection{Analytical interpretation}

The behavior of the density and the current obtained from the numerical solution is described by the analytic formula (\ref{cut}) (see fits in panels (b) and (c) of Fig.~\ref{fig1}).  Namely, the solution exhibits a cut singularity, indicative of weak solutions. The purpose of this section is to motivate this result.

\subsubsection{Bogolyubov equations and the necessity for weak solutions.}
 Here we present the perturbative solution of a pinned vortex in an otherwise uniform flow obtained from the zero-mode solution of the Bogolyubov equations. For this purpose it will be convenient to represent spatial position using polar coordinates $r,\phi$.

We consider a single vortex with $\sigma=1$ fixed at $r=0$ and set the
current to be fixed at large distances, i.e. $\vec{E}(r\to\infty)=-\he \v_c$.
Beyond the vortex core, representation \rref{BEGP} is valid.
Moreover, at the distances $v_c\ll r \ll 1/v_c$ the perturbation of
the vortex density and current profile is small and we can look for
a solution in the form
\be
\Psi(r,\phi)=e^{i\phi}\sqrt{n(r)}+ U(r)+V^*(r)e^{i2\phi},
\label{sol1}
\ee
The expected behavior of the solution at distances larger than the healing length,
\[
\Psi(r,\phi)=e^{i\phi+i\vec{v}_c\cdot \vec{r}}\sqrt{n(r)}
\simeq e^{i\phi}+\frac{iv_c r}{2} + \frac{iv_c^* r}{2} e^{i2\phi}, \quad v_c=v_c^x+iv_c^y,
\]
establishes the condition on amplitudes at $1\ll r \ll 1/|v_c|$:
\be
U^{(2)}(r)=-V^{(2)}(r)=i v_c r,
\label{cond1}
\ee
where the meaning of the superscript $(2)$ will become clear later on.
Substituting \req{sol1} into time independent \req{GP} and linearizing
with respect to $U$ and $V$, we obtain the zero-mode of the
Bogolyubov equations
\be
\begin{split}
&\left[-\frac{1}{2}\Dt+2n(r)-1\right]U(r)+ n(r)V(r)=0;\\
&\left[-\frac{1}{2}\Dt+\frac{2}{r^2}+2n(r)-1\right]V(r)+ n(r)U(r)=0.
\end{split}
\label{Bogolyubov1}
\ee
A convenient way to rewrite \req{Bogolyubov1} is to introduce even and odd
combinations $W_{\pm}=U\pm V$ for which we obtain
\be
\begin{split}
&\left[-\frac{1}{2}\Dt+3n(r)-1+\frac{1}{r^2}\right]W_+(r)- \frac{1}{r^2}W_-(r)=0;\\
&\left[-\frac{1}{2}\Dt+n(r)-1+\frac{1}{r^2}\right]W_-(r)- \frac{1}{r^2}W_+(r)=0.
\end{split}
\label{Bogolyubov10}
\ee
This form of the equations is useful for analysis of the solutions at large distances, $r\gg 1$.

Equations \rref{Bogolyubov1} form a coupled pair of second order
differential equations and therefore have four independent solutions.
One solution can be ruled out as it diverges exponentially at large
distances: $U(r)=V(r)=\exp(2r)$. Another solution can be also ruled
out since it contains a component which is too rapidly divergent at short distances:
$V(r)\simeq 1/r^2$.
The two remaining two solutions of \req{Bogolyubov1} are connected
by the invariance of the Wronskian:
\be
\frac{d}{dr}r\left\{
U^{(2)}\frac{d U^{(1)}}{dr}
+ V^{(2)}\frac{d V^{(1)}}{dr}
 - \left[(1)\leftrightarrow (2)\right]
\right\}=0.
\label{Wronski}
\ee

One (exact) solution of \req{Bogolyubov1} is easy to find
\be
\begin{pmatrix}
U^{(1)}
\\
V^{(1)}
\end{pmatrix}
= \begin{pmatrix}
\frac{d\sqrt{n}}{dr} + \frac{\sqrt{n}}{r}
\\
\frac{d\sqrt{n}}{dr} - \frac{\sqrt{n}}{r}
\end{pmatrix}
= \left\{\begin{matrix}\begin{pmatrix}
 1/{r}
\\ - {1}/{r}
\end{pmatrix} + {\cal O}\left(\frac{1}{r^3}\right), & r\gg 1;
\\[4mm]
\begin{pmatrix}
 2\alpha
\\ {\cal O}(r^2)
\end{pmatrix}, & r \ll 1.
\end{matrix}\right.
\label{strongsolution}
\ee
This solution can be understood as a small translation of the vortex as a whole, but it does not satisfy the boundary condition of supercurrent flow at large distances \rref{cond1}.

The second solution of \req{Bogolyubov1}, which can be obtained using the above solution and the invariance of the Wronskian (\ref{Wronski}), does  satisfy boundary conditions (\ref{cond1}) [the solution at $r\gg 1$ is obtained from \req{Bogolyubov10} with $W_+=W_-/(2r^2)+\dots, \n(r)=1-1/(2r^2)-1/(2r^4)- \dots$]:
\be
\begin{pmatrix}
U^{(2)}
\\
V^{(2)}
\end{pmatrix}
= \frac{iv_c}{2} \left\{\begin{matrix}\begin{pmatrix}
r+\frac{\ln r}{r}+\frac{1}{2r}\\ - r-\frac{\ln r}{r}+\frac{1}{2r}
\end{pmatrix} + {\cal O}\left(\frac{v_c \ln r}{r^3}\right), &1/|v_c| \gg r\gg 1
\\[4mm]
\begin{pmatrix}
 \frac{2}{\alpha}\ln r + {\rm const}
\\ {\cal O}(r^2)
\end{pmatrix}, & r \ll 1
\end{matrix}\right.
\label{weaksolution}
\ee
where the constant term is to be found from the consideration of the
short-distance physics. The solution \rref{weaksolution} is linear in $v_c$ and it diverges weakly for $r\to 0$. This type of divergent solution has been considered before in Ref. \cite{ThoulessAnglin2007}, however the short distance cutoff of the divergence was set there to be on order of the healing length. In the next subsection we study the
the fate of this divergence within the vortex core and the appearance of non-analytic behavior in $v_c$.

\subsubsection{Solution at small distances and relation to the Laplace and
 the Popov equations.}

\begin{figure}[h]
\includegraphics[width=1\columnwidth]{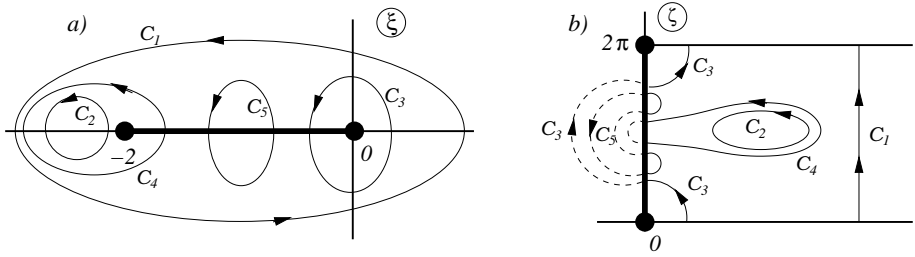}
\caption{The conformal mapping (\ref{elliptic1}) between the $\xi$-plane and a strip in the $\zeta$-plane whose lower edge should be identified with its upper edge. In the $\xi$-plane the vortex is located at the origin.
Contours ${\cal C}_{1,3}$ encompass the vortex. The other
contours do not enclose the vortex even though they may go through
the cut. Dashed lines in the $\zeta$-plane corresponds to paths on the second Riemann sheet in $\xi$-plane (not shown here).}
\label{elliptic}
\end{figure}

Let us now construct the solution of the Gross-Pitaevskii equations at distances much smaller
than the healing length. According to \reqs{sol1} and \rref{weaksolution}, at $r\ll 1$ this
solution can be sought for in a scaling form
\be
\Psi(z,z^*)=i\frac{v_c}{\alpha} \Upsilon\left(\frac{\alpha^2z}{i v_c};
  \frac{\alpha^2 z^*}{-i v_c^*} \right).
\label{scaling1}
\ee
At the distances $|v_c| \ll |z| \ll 1$, \req{scaling1} must reproduce
the solution of the Bogolyubov equation, i.e.
\begin{subequations}
\be
\Upsilon\left({\xi}; \xi^* \right)= \xi + \frac{1}{2}\ln \xi\xi^*;\quad |\xi| \gg 1.
\label{scaling2}
\ee
Moreover, for distances much shorter than the healing length, $|z|\ll 1$, the wave-function can be approximated by the solution of the Laplace equation
\be
\frac{\partial^2}{\partial \xi\partial \xi^*}\Upsilon\left({\xi}; \xi^* \right)=0,
\label{Laplace1}
\ee
everywhere, except for a set of measure zero. From the numerical
solution we know that this set should be a cut on the complex $\xi$-plane, see
Fig.~\ref{elliptic} (a), and current conservation imposes the boundary
conditions on the cut:
\be
\Upsilon\left({\xi}; \xi^* \right)=0; \quad \mathrm{Im}\,\xi=0;\ -2<
\mathrm{Re}\, \xi<0;
\label{Laplace2}
\ee
\label{Laplace}
\end{subequations}

Equations \rref{Laplace} are easily solved using elliptic coordinates
\be
\xi=\cosh{\zeta}-1; \quad  \xi^*=\cosh{\zeta^*}-1;
\label{elliptic1}
\ee
Transformation \rref{elliptic1} is the conformal mapping of the
complex $\xi$-plane to the strip $\mathrm{Re}\, \zeta \geq 0, \ 0\leq\mathrm{Im}\, \zeta < 2\pi$, see
Fig.~\ref{elliptic}. The cut corresponds to the line
$\mathrm{Re}\zeta=0$.

As the transformation \rref{elliptic1} is conformal, the Laplace equation (\ref{Laplace1}) holds its form also
for the new coordinates $\zeta,\zeta^*$, and therefore the desired solution
can be easily found:
\be
\Upsilon(\zeta,\zeta^*)=\frac{1}{2}\left[e^{\zeta}- e^{-\zeta^*} + \zeta +\zeta^*\right].
\label{solution2}
\ee
Notice that for $|\xi|\gg1$, $\zeta \simeq \ln \xi$ and the above solution reduces to (\ref{scaling2}), while on the line of purely imaginary $\zeta$ (which maps to the cut in the $\xi$-plane) the wave function vanishes (\ref{Laplace2}).

Equations \rref{solution2}, \rref{elliptic1}, and \rref{BEGP}, give
the complete solution of Popov's equations \rref{EQ:numerics} at
$\sigma=1,\ \vec{R}_0=0$ everywhere except on the cut singularity. The
excellent agreement with numerics is demonstrated in Fig.~\ref{fig1}

The remaining question to answer is how \req{solution2}, which is fixed on the
{\em cut}, reproduces the {\em point} source in the first of \reqs{eq:sGPE-2}. To
answer this question, we notice that according to \rref{BEGP},
in all regions where $|\Upsilon|>0$
\be
-  \vec{\nabla}\frac{\vec{\nabla} a_0}{\rho^2}=  \nnabla\frac{\vec{E}}{\rho^2}=-\vec{\nabla}\he \vec{\nabla}\theta;
\quad\theta(\zeta,\zeta^*)=\frac{i}{2}\ln\frac{\Upsilon^*(\zeta,\zeta^*)}{\Upsilon(\zeta,\zeta^*)}.
\label{E-phase}
\ee
This implies that  $\nabla\frac{\vec{\nabla} a_0}{\rho^2}=0$ is satisfied
outside the cut and only the cut region needs investigation, since on the
cut  $|\Upsilon|=0$, and the phase may experience an arbitrary jump. We will
define this jump in the original coordinates as
\be
\begin{split}
&\theta(\xi+i0)-\theta(\xi-i0)=
\theta[\zeta(\xi)]-\theta[2\pi i- \zeta(\xi)], \\ & \mathrm{Im}\,\xi=0;\ -2<
\mathrm{Re}\, \xi<0;
\end{split}
\label{overcut}
\ee
where here, in order to clarify the equation, we did not write explicitly the dependence of the phase on the conjugate variables. With the definition \rref{overcut} the evolution of the phase through
the cut on the $\xi$-plane, Fig.~\ref{elliptic} a) is equivalent to
the evolution along the dashed line on the $\zeta$-plane,  Fig.~\ref{elliptic} b).
Let us now calculate the two-dimensional integral of \req{E-phase}
over some region ${\cal A}_i$. We obtain
\be
\int_{{\cal A}_i} d^2r \vec{\nabla}\frac{\vec{E}}{\rho^2}=
\int_{{\cal C}_i}d\vec{\ell} \nnabla \theta
= \frac{1}{2}\int_{{\cal C}_i}\left[d\zeta\frac{\partial
    \theta}{\partial \zeta}+d\zeta^*\frac{\partial
    \theta}{\partial \zeta^*} \right].
\label{stokes}
\ee
Here ${\cal C}_i$ is the directed contour surrounding the area ${\cal
  A}_i$, and in the last equation we used the conformal transformation
\rref{elliptic1} and the definition of the phase jump \rref{overcut}.
Finally, using $\theta\left(\zeta+2\pi i\right)= \theta+2\pi$, we
obtain from inspection of Fig.~\ref{elliptic} b), that only the
contours going around point $\xi=0$ acquire the phase $2\pi$
\be
 \frac{1}{2}\int_{{\cal C}_i}\left[d\zeta\frac{\partial
    \theta}{\partial \zeta}+d\zeta^*\frac{\partial
    \theta}{\partial \zeta^*} \right]= 2\pi \int_{{\cal A}_i}  d^2 r \delta(\r).
\label{integral2}
\ee
As the area ${{\cal A}_i}$ is arbitrary, \reqs{integral2} and
\rref{stokes} reproduce the point source in \req{eq:sGPE-1}.

\subsection{Solution at large distances and the vortex dipole moment.}

The solution at distance larger than the healing length is obtained from Eqs. (\ref{strongsolution}--\ref{weaksolution}) in the range $1\ll r\ll 1/v_c$, by matching \reqs{scaling1}, \rref{solution2}, and \rref{elliptic1} with the linearized asymptotics  (\ref{strongsolution}--\ref{weaksolution}) at any $v_c \ll r\ll 1$. With logarithmic accuracy we find
\[
\begin{pmatrix}
U^{}
\\
V^{}
\end{pmatrix}
= \begin{pmatrix} U^{(2)}
\\
V^{(2)}\end{pmatrix}
+
\frac{iv_c}{2\alpha^2}\left[\ln\frac{1}{|v_c|}+{\cal O}(1)\right]\begin{pmatrix}
U^{(1)}
\\
V^{(1)}
\end{pmatrix}
\]
At $r\gg 1$, we find
\be
\begin{pmatrix}
U^{}
\\
V^{}
\end{pmatrix}
=
\frac{iv_c}{2} \begin{pmatrix}
r+\frac{\ln r}{r}+ \frac{ \ln(1/|v_c|)}{\alpha^2
  r}+\frac{1}{2r}\\[3mm] - r-\frac{\ln r}{r}-\frac{
  \ln(1/|v_c|)}{\alpha^2 r}+\frac{1}{2r}
\end{pmatrix} + {\cal O}\left(\frac{v_c \ln r}{r^3}\right),
\label{sol-dipole1}
\ee
Substituting \req{sol-dipole1} into \req{sol1} and the result in
\req{BEGP}, and recalling that $v=\he E/B$, we obtain Eqs. (\ref{dipole1}--\ref{dipole3})
of the introduction. This completes the derivation of
the microscopic expression of the dipole moment.

It is instructive to rewrite \req{dipole1} in Galilean invariant
form. Using \reqs{Galileantransform1}, taking into account the
relation between the superfluid velocity and current at large distances
$\vec{v}_c\simeq \he \vec{E}/B$ and returning to the two-dimensional
vector notation, we obtain for the field distribution around the
vortex $\vec{R}_i(t)$ with vorticity $\sigma_i$
\be
\begin{split}
&\frac{\vec{E}}{B}=\frac{\vec{E}_{i}}{B_i}+
\left[\frac{{\sigma_i}\vec{r}_i}{|\vec{r}_i|^2}+\frac{2(\d_i\cdot
    \vec{r}_i)\vec{r}_i-\vec{r}_i^{\, 2}\d_i}{|\vec{r}_i|^4}+{\cal
    O}\left(\frac{\vec{v}}{r_i^2}\right)\right];
\\
&
\vec{r}_i\equiv \vec{r}- \vec{R}_i(t);
\\
&
\d_i(r)=\f_i \ln |\vec{r}_i|+ \d(1); \quad
\d_i(1)= \frac{\f_i}{\alpha^2}\ln\frac{1}{|\f_i|},
\end{split}
\label{sol-dipole2}
\ee
where the force $\f_i$ was defined in \req{force1}, with the fields
understood as assuming their asymptotic limit far from the vortex (corresponding to $E_i$) .
It is important to emphasize that the dipole moment as well as the
force acting on vortex are Galilean invariant.

 Equations
\rref{sol-dipole2} are the basic ingredients for our construction of
an effective field theory describing physics over scales larger
than the healing length.

\section{Effective field theory at distances larger than the
  healing length.}
\label{sec:eft}
So far, we considered the solutions for a uniformly moving vortex in a
stationary flow. Now we wish to describe the full dynamics of the
vortices and fields in the limit of low frequency excitations. To do that we take advantage of  the
separation of length scales in the problem. Imagine an element of the
superfluid (which may or may not include a vortex) oscillating
at some frequency $\omega \ll 1$. The oscillatory nature of the motion of this element
can be observed only at distances larger or of the order of
$1/\omega$. At the smaller distances the velocity field is essentially uniform and
can be considered within the quasistatic approximation, for which the solution for a single vortex is known (see Sec.~3). This property allows  one to write
an effective theory in terms of slow fields which change over scales much larger than the
healing length, provided  we consider only situations where
the distances between vortices are much larger than the healing length.
(The interesting case of a vortex-antivortex pair at distances on order of or smaller than the healing length will be
considered elsewhere \cite{We1}).

Apparently, the most straightforward way to introduce the long-wavelength effective
theory is simply to drop the gradient $(\nnabla B)^2$ term in \req{action2}.
\be
{\cal S}_{eff}\overset{?}{=}
\int d^3 x  \left[ \frac{\vec{E}^2}{2B} - \frac{(B-1)^2}{2}-
2\pi \boldsymbol{J}^V \cdot \boldsymbol{A}\right].
\label{action?}
\ee
In the absence of vortex current, $\boldsymbol{J}^V=0$, this is a legitimate approximation.
However, when $\boldsymbol{J}^V \neq 0$ the above action diverges
 due to the $\delta$-functions in the vortex density and currents \rref{vortexcurrent}.  In other words, the
local hydrodynamical description is not applicable in the vicinity of the vortex core.

In order to regularize this divergence one has to replace the point like source by a smeared source of size
larger than the healing length $r_0\gtrsim 1$. There are many ways of smearing
the $\delta$-function. We find that the most convenient one, from the technical
point of view, is to replace the point-source by a ring-source of radius $r_0$, i.e.
\be
2\pi\delta(\vec{r})\to 2\pi\delta_{r_0}(\vec{r})
\equiv \frac{\delta\left(|\vec{r}|-r_0\right)}{r_0}.
\label{smeareddelta}
\ee
This form ensures that the perturbation of the density due to the contribution from the vicinity of the vortex can not be larger than $1/(2r_0^2)\lesssim 1$.

However, the effective theory (\ref{action?}) with the regularization (\ref{smeareddelta}) still does not account for the change in the density and the action due to the contribution from the core of the $l$-th vortex\footnote{The above regularization implies that fields inside the ring around the $l$-th vortex are due to the external fields and the fields created by all the vortices except $l$-th one.}.
However, this contribution is coming from
short distances, and can therefore be found for each vortex separately by employing the quasistationary approximation and using the results presented in Sec.~\ref{sec:solution} which provide a full description of the vortex core.

The resulting effective action takes the form
\be {\cal S}_{eff}= \int
d^3 x \left[ \frac{\vec{E}^2}{2B} - \frac{(B-1)^2}{2}- 2\pi
  \tilde{\boldsymbol{J}}^V \cdot \boldsymbol{A}\right] + \sum_l {\cal
  S}^V\left[\vec{R}_l(t), \f_l\right],
\label{action2-et}
\ee
where the vortex current $\tilde{\boldsymbol{J}}^V$ is calculated
using \req{vortexcurrent} with the regularization \rref{smeareddelta},
while the last term represents the contribution from the core of each
one of the vortices. This contribution is a functional of the vortex
velocity, $\partial_t \vec{R}_l$, and the force, $\vec{f}_l$, acting
on the vortex [see \req{force1}]:
\be
\begin{split}
& \f_l(t)=
\vec{E}_l+B_l\he\pt\vec{R}_l(t);
\\ 
&
\begin{pmatrix} \vec{E}_l\\B_l\end{pmatrix}
\equiv\int d^2r
\delta_{r_0^-}\left[\vec{r}-\vec{R}_l(t)\right]
\begin{pmatrix} \vec{E}(\vec{r},t)\\B(\vec{r},t)\end{pmatrix}.
\end{split}
\label{force2}
\ee
Here $\delta_{r_0^-}$ is the regularized $\delta$-function defined
in \req{smeareddelta}
 with the limit $r_0 \to r_0-0^+$ in accordance with the comment of footnote 1.

The form of the functional ${\cal S}^V\left[\vec{R}_l(t),
  \f_l\right]$ is invariant under translation, gauge, and Galilean
transformations. In the next section we shall show it has the form:
\be {\cal S}^V\left[\vec{R}(t), \f\, \right] =\int
dt \left[ \frac{ \pi }{\alpha^2}\vec{f}^{\, 2} \ln \frac{1}{|\vec{f}|}
  -\Delta E(\tilde{B}(\vec{R})) \right],
\label{Vortex-core-action}
\ee
where the constant $\alpha=0.8204\dots$ is defined in \req{n-small-z}.

The first term in the square brackets (dipole term)
is the non-analytic contribution due to the force $\f$
acting on the vortex which produces the cut singularity and the dipole
moment $\d$ discussed in
Sec.~\ref{sec:solution}. [Another way to represent the dipole term is
to write $\f\cdot \d(1)$, see \req{sol-dipole2}].

The second (energy) term, $\Delta E(\tilde{B}(\vec{R}(t))$ should be understood as the
contribution to the energy by the non-deformed core. One could think
that it is constant and can be neglected in the action, however,
this energy depends on the smooth part of the density (for example when $B\neq 1$ due to wave deformation) and will contribute in the
resulting equation of motion.
The ``smoothened'' field $\tilde{B}(\vec{R})$ can be defined as an average
of the effective field $B$ inside the ring
\be
\tilde{B}(\vec{R})=\frac{1}{\pi r_0^2}\int d^2r
\Theta(r_0-|\vec{r}-\vec{R}|){B}(\vec{r}),
\label{Btilde}
\ee
where $\Theta(x)$ is the Heaviside step function.

In principle, it is possible to have an additional contribution allowed by Galilean
invariance: The term $(\partial_t \vec{R})^2$, which adds only total time derivatives to the Lagrangian under Galilean transformations. [This kinetic term can be interpreted as an additional contribution to the
kinetic energy due to the fact that the true density in the vortex is
lower than the one obtained by the effective theory].
However, we shall show that this term is absent within our consistent
effective theory, see Secs.~\ref{sec:et-core} and  \ref{sec:cm1}.

To conclude the description of the effective vortex we mention the
relation of the dipole term $\f\cdot \d(1)$ to the calculation of the
logarithmically divergent mass of the vortex
\cite{Feynmann55,Popov73, DuanLeggett92,ThoulessAnglin2007}. The latter approach consists of writing $\d=\f \ln
|\vec{r}_0|$ and considering $r_0$ to be very large (on order of
the intervortex distance) to
include the whole dipole deformation of the suprefluid field in the vortex
mass, thus effectively decoupling the superfluid waves and the vortex
motion. Our effective theory
defined on the scale $r_0\gtrsim 1$
includes all these contribution but treats them on the level of the
 deformation of the superfluid fields $\cal B,{\cal E}$ described by the
 equation of motion. However, our theory
includes the previously overlooked non-analytic contribution  $\d(1)\sim \f\ln(1/|f|)$ coming
from the core of the vortex. We will see later that this
non-analytic contribution is decisive for the slow vortex
dynamics, see Sec.~\ref{sec:cm} for more details.

\subsection{Derivation of the effective action for the vortex core.}
\label{sec:et-core}

In this section we derive formula (\ref{Vortex-core-action})
describing the contribution to the action from the vicinity of the
vortex. It is clear by construction of the effective action
\req{action2-et} that in order to calculate ${\cal S}^V$ it is sufficient
to consider a single vortex, characterized by its coordinate
$\vec{R}(t)$ and its vorticity $\sigma$.

The action ${\cal S}^V$  can be understood as the difference between the exact action,
${\cal S}_{\cal B,E}$, and the effective action of the smooth fields,
${\cal S}_{ B,E}$, where both contributions are evaluated within a
circle of radius $r_0$ around the vortex position $\vec{R}$:
\begin{subequations}
\label{et-derivation-0}
\be
{\cal S}^V={\cal S}_{\cal B,E}-{\cal S}_{ B,E}.
\label{et-derivation-0a}
\ee
The exact
action can be found by using the fields, $\vec{\cal E}, \ \vec{\cal
  B}$, describing the quasistationary solution of the equations of
motion, while the effective action is the long-wavelength action described
by the integral on the right-hand side of \rref{action2-et} (we reserve the $\vec{ E}, \ B$ notation for the smooth fields introduced in this
action).

The subtraction of ${\cal S}_{ B,E}$ is designed to simply exclude the
contribution of the smooth fields near the vortex core where the local
description of the action is not applicable,
\be
\begin{split}
{\cal S}_{ B,E}&=
\int_{|\vec{r}-\vec{R}(t)|<r_0}d^3x
\left[\frac{\vec{E}^2}{2B}
- \frac{\left({ B}-1\right)^2}{2}
\right]
\\
&
- 2\pi\sigma \int d^3x\delta_{r_0}\left[\vec{r}-\vec{R}(t)\right]
\left[a_0(\vec{r},t)-\vec{a}(\vec{r},t)\pt \vec{R}\right].
\end{split}
\label{et-derivation-0b}
\ee
where in the last term we used explicit expressions \rref{vortexcurrent}
and \rref{smeareddelta} for the vortex density current
$\tilde{\boldsymbol{J}}^V$.

To evaluate the contribution ${\cal S}_{\cal B,E}$, it is convenient to use a coordinate system which is attached to the vortex position, $\vec{R}(t)$. Thus all the fields should be transformed to
the moving (not necessarily inertial) frame according to \req{Galileantransform1} with
$\vec{v}\to \pt\vec{R};\ \vec{v}t\to  \vec{R}(t)$:
\be
\begin{split}
{\cal S}_{\cal B,E}&=\int_{|\vec{r}|<r_0}d^3x
\left\{
\frac{\left(\vec{\cal E}-{\cal B}\he \pt \vec{R} \right)^2}{2{\cal B}}
-
\left[\frac{\left({\cal B}-1\right)^2}{2}
+ \frac{\left(\vec{\nabla}\sqrt{\cal B}\right)^2}{2}
\right]
\right\}
\\
&
- 2\pi \int_{|\vec{r}|<r_0} d^3x\, \tilde{a}_0\, J_0^V.
\end{split}
\label{et-derivation-0c}
\ee
Galilean invariance \rref{Galileantransform1} and the
continuity of the fields on the ring of radius $r_0$ around the vortex imply that
\be
\begin{split}
&\tilde{a}_0(\r,t)\Bigg|_{|\r|=r_0}
=\left\{a_0[\r+\vec{R}(t),t]-\vec{a}[\r+\vec{R}(t),t]\pt\vec{R}(t)\right\}\Bigg|_{|\r|=r_0};
\\
&
\tilde{\vec{a}}(\r,t)\Bigg|_{|\r|=r_0}=\vec{a}[\r+\vec{R}(t),t]\Bigg|_{|\r|=r_0};
\\
&
 \frac{\vec{\cal E}-{\cal B}\he \pt \vec{R}}
{\cal B}
\Bigg|_{|\r|=r_0}
 =\frac{\vec{E}}{B}\Bigg|_{|\r|=r_0-0^+}
 + \left[\frac{{\sigma}\vec{r}}{|\vec{r}|^2}+\frac{2(\d\cdot
     \vec{r})\, \vec{r}-\vec{r}^{\, 2}\d}{|\vec{r}|^4}+\dots\right]\Bigg|_{|\r|=r_0},
\end{split}
\raisetag{60pt}
\label{et-derivation-0d}
\ee
where $\tilde{a}_0(\r,t)$ and $\tilde{\vec{a}}(\r,t)$ denote the smooth gauge fields of the effective action, and we use the definition $|\r|=r_0-0^+$ for the smooth fields (see footnote 1).
The last equation is obtained using the dipole expansion for the field outside the vortex \rref{sol-dipole2}.
\end{subequations}

We now turn to calculate of the contribution ${\cal S}_{\cal B,E}$, which is a functional of the quasi-static fields.
Consider, first, the source term in (the second line of) \rref{et-derivation-0c}. Using the equation of motion \rref{popov11b}, and a Galilean transformation to the coordinate system comoving with the vortex \rref{Galileantransform1}, we can rewrite it as
\[
- 2\pi \int_{|\vec{r}|<r_0} d^3x\, \tilde{a}_0\, J_0^V =
-\int_{|\vec{r}|<r_0}d^3x\tilde{a}_0\vec{\nabla}\cdot\left(\frac{\vec{\cal E}-{\cal B}\he \pt \vec{R}}{\cal B}\right).
\]
Integrating by parts and utilizing the connection between the fields and vector potential \rref{fields}, we obtain
\be
\begin{split}
- 2\pi \int_{|\vec{r}|<r_0} d^3x\, \tilde{a}_0\, J_0^V& =
-\int
dt\ointctrclockwise_{|\vec{r}|=r_0}\tilde{a}_0\left[\left(\frac{\vec{\cal E}-{\cal B}\he \pt \vec{R} }{\cal B}\right)\he d\vec{r}\right]
\\
&
-\int_{|\vec{r}|<r_0}d^3x\left(\frac{\vec{\cal E}-
{\cal B}\he \pt \vec{R} }{\cal B}\right)
\left(\vec{\cal E}+\pt \vec{\tilde{a}}\right).
\end{split}
\label{et-derivation-1}
\ee
Here, and henceforth, the vector $d\vec{r}$ appearing in the surface terms, is directed along the contour.
We now eliminate the time derivative of the vector potential via integration by parts and the equation of motion \rref{popov11c}:
\be
\pt\left(\frac{\vec{\cal E}-{\cal B}\he \pt \vec{R} }{\cal B}\right)=
\he\nnabla\cH;\quad \cH\equiv \left(
 \frac{\vec{\cal E}^2}{2{\cal B}^2}+B-1-\frac{1}{2\sqrt{\cal
     B}}\nnabla^2\sqrt{\cal B}
\right).
\label{et-Maxwell-1}
\ee
This reduces \req{et-derivation-1} to
\be
\begin{split}
 - 2\pi& \int_{|\vec{r}|<r_0} d^3x\, \tilde{a}_0\, J_0^V =
 -\int
 dt\ointctrclockwise_{|\vec{r}|=r_0}\tilde{a}_0\left[\left(\frac{\vec{\cal E}-{\cal B}\he \pt \vec{R} }{\cal B}\right)\he d\vec{r}\right]
 \\
 &
 -\int_{|\vec{r}|<r_0}d^3x\left[\left(\frac{\vec{\cal E}-{\cal B}\he \pt \vec{R} }{\cal B}\right)\vec{\cal E}
 -
 \vec{\tilde{a}}\he\vec{\nabla} \cH
 \right].
\end{split}
\label{et-derivation-2}
\ee
Integrating the last term by parts and using $\vec{\nabla}\he\vec{\tilde{a}}={\cal B}$, we arrive at an expression which separates the contribution from the interior of the vortex core [denoted below by $(b)$], from the surface terms coming from the interface between the core the exterior of the vortex [denoted by $(s)$]:
\be
\begin{split}
& - 2\pi \int_{|\vec{r}|<r_0} d^3x\, \tilde{a}_0\, J_0^V
=\Sigma_{b}+\Sigma_{s};
\\
& \Sigma_{b}= \int_{|\vec{r}|<r_0}d^3x
\left[-\left(\frac{\vec{\cal E}-{\cal B}\he \pt \vec{R} }{\cal B}\right)\vec{\cal E}+ {\cal B}\cH \right];
\\
&  \Sigma_{s} = - \int dt \ointctrclockwise_{|\vec{r}|=r_0}
\left[
 \frac{\tilde{a}_0\left(\vec{\cal E}-{\cal B}\he \pt \vec{R} \right)\he }{\cal B}
+
\cH
\tilde{\vec{a}}
\right]d\vec{r}.
\end{split}
\raisetag{80pt}
\label{et-derivation-3}
\ee
The surface contribution $\Sigma_{s}$ can be now evaluated using the
asymptotic expression of the fields at $r=r_0$, see
\req{et-derivation-0d}. With the same accuracy of this asymptotic expansion we find from \req{et-Maxwell-1}
\be
{\cH}\Bigg|_{|\vec{r}|=r_0}=
-\partial_t\left(\frac{\vec{r}\he \vec{E}}{B}\right)\Bigg|_{|\vec{r}|=r_0-0^+} - \partial_t\left[ \frac{\vec{r} \he \d(t,r_0)}{r^2}\right]\Bigg|_{|\vec{r}|=r_0}.
\label{et-derivation-4}
\ee
We substitute \reqs{et-derivation-0d} and \rref{et-derivation-4} into
\req{et-derivation-3}, and use the following relations satisfied by an arbitrary smooth function $b(\vec{r})$
\be
\begin{split}
&\ointctrclockwise_{|r_1|=r_0} \frac{\vec{r}_1\he d\vec{r}_1}{|\vec{r_1}|^{\, 2}}b(\vec{r}+\vec{r}_1)
= \oint_{|r|_1=r_0}  \frac{dr_1}{r_1}b(\vec{r}+\vec{r}_1)=2\pi \int d^2r_1\delta_{r_0}(\vec{r}-\vec{r}_1)b(\vec{r}_1);
\\
&\ointctrclockwise_{|r_1|=r_0} \frac{\left[2\vec{r}_1\left(\d\cdot\vec{r}_1\right) - \d\, \vec{r}_1^{\, 2}
\right]\he d\vec{r}_1}{|\vec{r_1}|^{\, 4}}b(\vec{r}+\vec{r}_1)\simeq
\pi \int d^2r_1\delta_{r_0}(\vec{r}-\vec{r}_1)\, \d\cdot\vec{\nabla} b(\vec{r}_1);
\end{split}
\label{et-derivation-5}
\ee
where the smeared $\delta$-function is defined in \req{smeareddelta}.
We obtain
\be
\begin{split}
\Sigma_s&=\int d^3x\delta_{r_0^-}\left[\vec{r}-R(t)\right]
\\
&\times
\left\{\pi r_0^2\left[\frac{\vec{E}^2}{2B}\right]_{t,\vec{r}}
-2\pi\sigma \left[a_0-\vec{a}\pt \vec{R}\right]_{t,\vec{r}}
+\pi\d\cdot\left[\vec{E}+B\he\pt\vec{R}\right]_{t,\vec{r}}\right\},
\end{split}
\label{et-derivation-6}
\ee
and the smeared $\delta$-function, $\delta_{r_0^-}(\vec{r})$ is defined by \req{smeareddelta} with $r_0\to r_0-0^+$.

We complete the derivation by substituting \req{et-derivation-6} into
\req{et-derivation-3} and plugging the result into
\reqs{et-derivation-0}. Performing analogous transformations,
neglecting the boundary term $\vec{r}\vec{\nabla} B$ at $|\vec{r}|=r_0$,
and using the expression \rref{dipole2} for the dipole moment,
we obtain\footnote{With the desired accuracy $\d(r_0)=\d(1)$.}
\begin{subequations} \label{action-final}
\be
{\cal S}^V=\int {dt}\left[{\cal L}_{d}+{\cal L}_K - \Delta E\right]
\label{action-final-a}
\ee
The dipole Lagrangian is the main result of this section as it describes a contribution which is non-analytic in the applied force,
\be
{\cal L}_d=\frac{\pi\f^{\, 2}\ln \frac{1}{|\f\, |}}{\alpha^2},
\label{action-final-b}
\ee
where the force acting on vortex was defined in \req{force2}.
Needless to say the dipole Lagrangian is Galilean invariant as the force acting on vortex  $2\pi\sigma\f$ is Galilean invariant.

The term $\Delta E$ is simply the contribution of the non-deformed vortex core to the energy:
\be
\Delta E=\int_{|\vec{r}-\vec{R}(t)|<r_0}d^2r\left[\frac{1-{\cal B}^2 }{2}-\frac{\left(B-1\right)^2}{2}\right],
\label{action-final-c}
\ee

Finally, the term ${\cal L}_K$ describes the analytic velocity dependence of the energy of the vortex.
Its formal expression is
\[
{\cal L}_K=\int_{|\vec{r}-\vec{R}(t)|<r_0}\left[\frac{{\cal B}\left(\pt\vec{R}\right)^2}{2}-\frac{\vec{E}^{\, 2}}{2B}
+ \int d^2r_1 \delta_{r_0^-}\left(\vec{r}_1-\vec{R}\right)\left(\frac{\vec{E}^{\, 2}}{2B}\right)_{\vec{r}_1}
\right];
\]
For homogeneous fields the last two terms cancel each other, so to explicitly calculate this term would require us to investigate the internal structure of the moving vortex within the effective theory. Such study is, however, not necessary as the form of ${\cal L}_K$ can be deduced from the Galilean invariance. The latter dictates that ${\cal L}_K$ can include only combinations of $(\pt R)^2$ (as it adds only total a time derivative to the Lagrangian under Galilean transformations) and any analytic function of $\f^{\, 2}$. However, an analytic function of $\f^{\, 2}$ is negligible in comparison with non-analytic dipole Lagrangian \rref{action-final-b}, and therefore
\be
{\cal L}_K=\beta\left(\pt \vec{R}\right)^2,
\label{action-final-d}
\ee
where $\beta$ is a coefficient of order one which can be found using Galilean invariance once again. Indeed, starting with a system of vortices at rest and switching to a coordinate frame moving with the velocity $\vec{v}$, dipole moments in this case do not arise and the terms \rref{action-final-b} and \rref{action-final-c} do not change the action. On the other hand, the original action transforms according to \req{ActionGalilean}, i.e. the coefficient in \req{action-final-d} is just the difference between the number of particles in the effective stationary vortex core and true stationary vortex core:
\be
\beta=\int_{|\vec{r}-\vec{R}|<r_0}d^2r\left[{\cal B}-B\right].
\label{action-final-e}
\ee
\end{subequations}
We will show in Sec.~\ref{sec:cm1}, that the density profile in
effective theory is arranged to exactly reproduce the same number of particles
in the vortex core as the microscopic solution, so that $\beta=0$.
The action \rref{Vortex-core-action} follows from \reqs{action-final}.

\subsection{The equations of motion.}
\label{sec:cl-em}

The derivation of the equations of motion resulting from Eqs.~(\ref{action2-et}--\ref{Btilde}) is analogous  to the derivation of
\reqs{popov11}. Because \req{action2-et} is a gauge theory, the first
Maxwell equation is still valid,
\begin{subequations}
\label{et-popov11}
\be
\partial_t B=-\nnabla\times \vec{E}.
\label{et-popov11a}
\ee
Instead of \req{popov11b} we obtain
 \be
\nnabla \cdot \left(\frac{\vec{E}}{B}\right) =2\pi
\tilde{J}^V_0-\vec{\nabla} \cdot \vec{\cal P}_V,
\label{et-popov11b}
\ee
where the smoothened vortex density-current is calculated
using \req{vortexcurrent} with the regularization
\rref{smeareddelta}, corresponding to a simple smoothening of the
original action. The last term, new in comparison with \req{popov11b},
is the change in vortex density due to polarization of the
vortex shape by an external force:
\be
 \vec{{\cal P}}_V
 =\sum_l \delta_{r_0^-}[\vec{r}-\vec{R}_l(t)]\vec{d}_l; \quad
\vec{d}_l=\frac{\partial}{\partial f_l}
\left(\frac{\pi}{\alpha^2}\f^{\, 2}_l\ln\frac{1}{|\f_l|}\right),
\label{et-popov11c}
\ee
see \reqs{force2}, \rref{smeareddelta}  for definitions of the force $\f_l$ and the smoothen
function $\delta_{r_0^-}$. Equation \rref{et-popov11b} is analogous to
the generalization of electrodynamics for polarizable media.

Equation \rref{popov11c} is replaced by
\be
\he\nnabla \left[\frac{1}{2}\left(\frac{\vec{E}}{B}\right)^2+B\right]=
 \he \nnabla {\cal
   M}+\vec{j}_v+\partial_t\left(\frac{\vec{E}}{B}+\vec{\cal P}_v\right)
\label{et-popov11d}
\ee
i.e. polarization and magnetization currents are now added to the vortex current.

The magnetic moment density is the derivative
of the effective action \rref{Vortex-core-action} with respect to
local magnetic field. We present it as the sum of two parts
\be
{\cal M}=\sum_l\left\{\delta_{r_0^-}[\vec{r}-\vec{R}_l(t)]\mathfrak{m}_l^{(d)}
+ \frac{\Theta(r_0-|\vec{r}-\vec{R}_l|)}{\pi r_0^2}\mathfrak{m}_l^{(c)}
\right\}.
\label{et-popov11e}
\ee
The first term is the magnetic moment arising from the motion of the
cut singularity:
\be
\mathfrak{m}_l^{(d)}=-\frac{\partial}{\partial
  B_l}\frac{\pi}{\alpha^2}\f^{\, 2}_l\ln\frac{1}{|\f_l|},
\label{et-popov11f}
\ee
the second term is the magnetization due to the circular currents in
the non-perturbed core:
\be
\mathfrak{m}_l^{(c)}
=-\frac{\partial \Delta E(B_l)}{\partial B_l}.
\label{et-popov11g}
\ee

\end{subequations}

Finally, we derive an equation of motion for the vortex. Varying
Eqs. (\ref{action2-et}--\ref{Vortex-core-action}) with respect to the
vortex coordinate $\vec{R}_l(t)$ we obtain instead of \req{force1}
\be
\frac{d}{d t}\frac{\partial}{\partial (\partial_t\vec{R}_l)}
\left(\frac{\pi}{\alpha^2}\f^{\, 2}_l\ln\frac{1}{|\f_l|}\right)
= 2\pi\sigma_l \f_l+ \frac{\partial}{\partial \vec{R}_l}
\left(\frac{\pi}{\alpha^2}\f^{\, 2}_l\ln\frac{1}{|\f_l|}\right)
+\mathfrak{m}_l^{(c)}\frac{\partial B_l}{\partial \vec{R}_l}.
\label{et-vortex-motion}
\ee
As opposed to \req{force1}, equation (\ref{et-vortex-motion}) describes the
complete vortex dynamics with second time derivatives generated by
the interaction with the phonon field.

It is worth emphasizing that \reqs{et-popov11} and
\rref{et-vortex-motion} are Galilean invariant by construction.
In the next section we will solve the classical equations of motion
for the most important physical situations.

\section{Solutions of the classical equations of motion.}

\label{sec:cm}

\subsection{The effective theory of a single vortex and calculation of
  constants in the effective action.}
\label{sec:cm1}

\begin{figure}[h]
\begin{center}
\includegraphics[width=0.7\columnwidth]{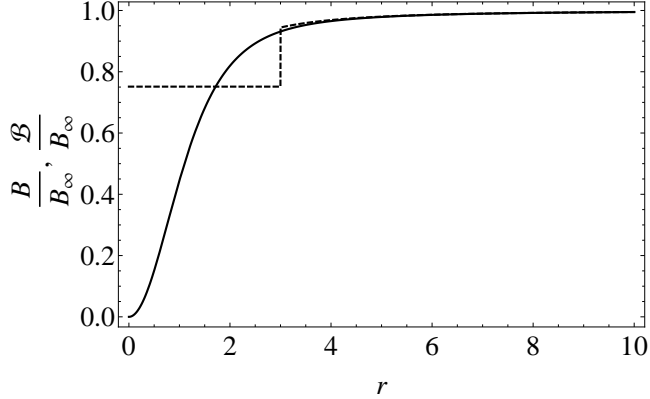}
\caption{Comparing of the particle density $B$ in the effective theory (dashed line) and the exact form (solid line)}
\label{fig:effective}
\end{center}
\end{figure}

Let us begin with the consideration of a free vortex located at the origin, $\vec{R}=0$, in the framework of the effective equations of motion \rref{et-popov11}.
The purpose of this study is to compare the resulting vortex shape
with the exact one and find the expressions for the core energy
$\Delta E$ and for the coefficient $\beta$ of \req{action-final-e}.
As the problem is axially symmetric, $\vec{\nabla}\he\vec{E}=0$ automatically, $\f=0$,
and \req{et-popov11b} takes the form
\[
\nnabla \cdot \left(\frac{\vec{E}}{B}\right)=\frac{\sigma}{r_0}\delta\left(|\vec{r}|-r_0\right),
\]
with the obvious solution
\be
\frac{\vec{E}}{B}=\frac{\sigma\vec{r}}{|\vec{r}|^2}\Theta\left(|\vec{r}|-r_0\right),
\label{cm-popov11b}
\ee
where $\Theta(x)$ is the step function.
Integrating \req{et-popov11d} to obtain the Bernoulli form,
and substituting \req{cm-popov11b} into the result, we find the magnetic field, $B(r)$, of the effective theory
\be
B(r)=B_\infty -\frac{\Theta(r-r_0)}{2r^2}-\frac{\Theta(r_0-r)}{\pi r_0^2}\frac{\partial \Delta E}{\partial B_\infty}.
\label{cm-Bv1}
\ee
 Thus effective theory reproduces correctly the leading asymptotic behavior in  the exterior domain $r>r_0$  (up to integrable terms). In the interior domain, $r<r_0$,
the effective theory may either overestimate or underestimate the density,
see Fig.~\ref{fig:effective}. This difference will manifest itself in the value of the coefficient $\beta$, as follows from \req{action-final-e}. Using $B(r)$ from \req{cm-Bv1} and utilizing the axial symmetry we find
\be
\beta=2\pi\int_0^{r_0}dr \, r\left[{\cal B}-B_\infty\right]+\frac{\partial \Delta E}{\partial B_\infty},
\label{cm-beta-1}
\ee

The energy of the core $\Delta E$ with the condition $B(r \gg 1)\to
B_\infty$, and the field inside the vortex ${\cal B}$ can be found from the
minimum of the energy functional
\be
\begin{split}
&\frac{\delta\Delta E\left[B_{\infty},{\cal  B}(r)\right]}{\delta {\cal
      B}(r) }=0;\\
&\Delta E\left[B_{\infty},{\cal  B}(r)\right]=
2\pi \int^{r_0}_0  drr\left[
\frac{\cal B}{2r^2}+\frac{1}{2}\left(\frac{\sqrt{\cal B}}{dr}\right)^2+\frac{\left({\cal B}-B_\infty\right)^2}{2}
\right].
\end{split}
\label{cm-beta-2}
\ee
From here it follows that
\[
\frac{\partial \Delta E}{\partial B_\infty}=2\pi 
\int_0^{r_0} drr\left(B_\infty-{\cal B}\right),
\]
and substituting in \req{cm-beta-1}, we obtain
\be
\beta=0.
\label{cm-beta-3}
\ee

This result implies that our effective theory not only
correctly describes the density $B$ outside the vortex core but also
conserves the number of particle inside the core of the
vortex. Therefore, the Galilean invariance is preserved automatically
and there is no need for the extra term $(\pt \vec{R})^2$ in order to preserve Galilean
invariance \rref{ActionGalilean}.

\subsection{Classical oscillatory motion of the vortex in the frozen background
  approximation}
\label{sec:cm2}

The frozen background approximation corresponds to fixing $B=B(t)$, and
$\vec{E}=\vec{E}(t)$ to be explicitly independent of the vortex position, $\vec{R}(t)$.
In this case \req{et-vortex-motion} reduces to
\be
\frac{d}{d t}\frac{\partial}{\partial\f}
\left(\frac{\f^2}{2\alpha^2}\ln\frac{1}{|\f|}\right)
+ \he\f=0;\quad
\f=  \left[\vec{E}(t)+ B(t)\he \pt\vec{R}(t)\right];
\label{m1}
\ee
where without loss of generality we also choose $\sigma=1$.

Equation \rref{m1} describes the motion of the vortex decoupled from
the rest of the superfluid in a very particular way: If some flow $\vec{E}(t)$, constant
in space, is applied to the system, the vortex is
displaced to a new equilibrium position such that the force $2\pi\f$ acting
on the vortex remains intact. In this sense the dynamics of the
force is decoupled from the rest of the superfluid,
(for the origin of the residual coupling and its physical consequences
see
Secs.~\ref{sec:cm4} and \ref{sec:cm5}).

Equation \rref{m1} has an obvious solution
\be
f_+= f_x+if_y=Fe^{{i\Omega_c t}}; \quad f_-\equiv f_x-if_y=F^*e^{-{i\Omega_c t}},
\label{m2}
\ee
where the effective ``cyclotron frequency'' is given by
\be
\Omega_c(|F|) =\frac{\alpha^2}
{\ln \frac{1}{|F|}-\frac{1}{2}} \ll 1.
\label{m3}
\ee

Equation \rref{m3} is the main result of this section. Indeed,
previous studies, see {\em e.g.} Refs.~\cite{Feynmann55,Popov73,DuanLeggett92,ThoulessAnglin2007}, indicated that allowing for a vortex mass in the equation of motion (thereby
allowing for a non-vanishing force acting on the vortex) may produce only
oscillations with frequency of the order of the healing frequency
$\hbar/(m\xi^2)$ (one in our units) which is beyond applicability of the local
hydrodynamics. However, from \req{m3} it follows that the non-analytic
reconstruction of the vortex core produces an oscillation frequency which is much smaller than
one, and therefore relevant also for the phonon dynamics. Moreover, the
oscillation frequency  $\Omega_c(|F|)$ depends in a non-linear manner on the
amplitude of the oscillation $F=\Omega_cR$ and vanishes
when $R\to 0$. Thus the lower limit on oscillation frequency is governed only by
quantum fluctuations, See Sec.~\ref{sec:qm}.

\subsection{Coupling the vortex motion to phonons}
\label{sec:cm3}
In the remainder of this section we will study the residual coupling
of a moving vortex to sound waves (phonons) in the superfluid.
The main effects are: (i) emission of sound waves by the
oscillating vortex, see \req{m2}, leading to cyclotron radiation of
phonons and to a radiation reaction force on the vortex; and (ii)
elastic scattering of phonons by the oscillating vortex.

Both effects are most conveniently studied in the basis of
the eigenfunctions of the Popov equations \rref{et-popov11}
linearized with respect to the solution of the motionless vortex
\rref{cm-popov11b} -- \rref{cm-Bv1}. We discuss partial waves solutions of
this linearized equation and the corresponding scattering phases are
discussed in subsection \ref{sec:cm30}. Then, in the following subsections, we employ the partial wave expansion to study the radiation and elastic scattering problems.

\subsubsection{Partial wave expansion outside and inside the effective
  vortex core}
\label{sec:cm30}

We linearize  Eqs.~(\ref{cm-popov11b}-- \ref{cm-Bv1}) with respect to the motionless vortex solution, fixing it at $\vec{R}=0$. For $|\vec{r}| \neq r_0$ any solution of \req{et-popov11b} can be parameterized
as
\be
\frac{\vec{E}}{B}=
\frac{\sigma\vec{r}}{|\vec{r}|^2}\Theta\left(|\vec{r}|-r_0\right)
-\he\vec{\nabla}\theta(\vec{r},t),
\label{pw1}
\ee
where the potential $\theta(\vec{r},t)$ is to be found from the other Popov
equations. Substituting \req{pw1} into \req{et-popov11d} and linearizing with
respect to $\vec{\nabla}\theta$ we find for the exterior domain of the effective vortex core,
$|\vec{r}|>r_0$,
\begin{subequations}
\label{pw2}
\be
B(\vec{r},t)=1-\frac{1}{2r^2}+\frac{\sigma\vec{r}}{|\vec{r}|^2}\he\vec{\nabla}\theta
- \pt\theta.
\label{pw2a}
\ee
This equation replaces the Bernoulli equation for the non-stationary case. In the interior domain, $|\vec{r}|<r_0$,  the same procedure yields
\be
\label{pw2b}
B(\vec{r},t)=1-\frac{1}{\pi r_0^2}\frac{\partial \Delta E}{\partial B_\infty}\Bigg|_{B_\infty=1}-\pt\theta,
\ee
\end{subequations}
instead of \req{cm-Bv1}.

Equations \rref{pw1} and \rref{pw2} enable us to find the
electric field of the vortex in the framework of the effective theory. Keeping only terms linear in $\vec{\nabla}\theta$ and $\pt\theta$, we obtain
\begin{subequations}
\label{pw3}
\be
\vec{E}=\frac{\sigma\vec{r}}{r^2}\left[1-\frac{1}{2r^2}\right] - \frac{\sigma\vec{r}}{r^2}\pt\theta
-\he\nnabla\theta+
\frac{\vec{r}^{\, 2} +
2\vec{r}\otimes\vec{r}
}{2|\vec{r}|^4}\,
\he\nnabla\theta.
\label{pw3a}
\ee
for $|\vec{r}|>r_0$. The last term in this equation
will be of no importance for future manipulations and we bring it here
for completeness only. Similarly, in the region $|\vec{r}|<r_0$ we find
\be
\vec{E}=-\left[1-\frac{1}{\pi r_0^2}\frac{\partial \Delta E}{\partial B_\infty}\Bigg|_{B_\infty=1}\right]\he\nnabla\theta.
\label{pw3b}
\ee
\end{subequations}

The equation for the potential, $\theta$, is now obtained by
substituting \reqs{pw2} and \rref{pw3} into the first Maxwell equation
\rref{et-popov11a}.  In the exterior domain, $|\vec{r}|>r_0$, it yields
\begin{subequations}
\label{pw4}
\be
\left[\nnabla^2-\pt^2 + \frac{2\sigma\vec{r}}{|\vec{r}|^2}\he\vec{\nabla}\pt\right]\theta
=-\he\nnabla\frac{\vec{r}^{\, 2} +
2\vec{r}\otimes\vec{r}
}{2|\vec{r}|^4}\,
\he\nnabla\theta.
\label{pw4a}
\ee
One can see, by power counting, that for  $|\vec{r}|\gg 1$ the right hand side of \req{pw4a} is not important and can be neglected.

Within the interior domain, $|\vec{r}|<r_0$, we make use of our working assumption that all relevant frequencies are much smaller than one. This allows us to write the the first Maxwell equation in the quasistatic approximation, i.e
\be
\nnabla^2\theta=0.
\label{pw4b}
\ee
\end{subequations}

Equations \rref{pw4} are axially symmetric and can be decomposed into
partial waves
\be
\theta(\vec{r},t)=\sum_{\nu,\omega}e^{-i\omega t}\left(\frac{x+i\sigma y}{r}\right)^\nu\theta_{\omega,\nu}(r)
;\quad
\theta_{\omega,\nu}=\theta_{-\omega,-\nu}^*.
\label{pw5}
\ee

\begin{subequations}
\label{pw6}
\noindent Then, \reqs{pw4} acquire the form of the Bessel equation
\be
\left[-\Dt+\frac{\nu^2-2\nu\omega}{r^2}+\omega^2\right]\theta_{\omega,\nu}=0 ~~~~~\mbox{for}~~~~  |\vec{r}|>r_0,
\label{pw6a}
\ee
while
\be
\theta_{\omega,\nu} \propto r^{|\nu|}~~~~~\mbox{for}~~~~ |\vec{r}|<r_0.
\label{pw6b}
\ee
\end{subequations}

The choice of the particular form of the solution of \req{pw6a} depends on the problem at
hand (here we assume that $\omega>0$). If one is interested in the problem
of the cyclotron radiation by the rotating vortex, the relevant choice should have
outgoing wave asymptotics {\em i.e.}
\be
\theta_{\omega,\nu}(r)\propto H^{(1)}_{|\nu|-\omega\sgn \nu}(\omega r),
\label{pw7}
\ee
where $H^{(1)}_n$, is the Hankel function of $n$-th order of the first kind.
(We use $\omega\ll 1$ for the calculation of the order of the
Hankel function). On the other hand, if one seeks to describe the elastic scattering of phonons from a vortex, the relevant solution has the form
\be
\theta_{\omega,\nu}(r)\propto \left[J_{|\nu|-\omega\sgn \nu}(\omega r)\cos\tilde{\gamma}_\nu -Y_{|\nu|-\omega\sgn \nu}(\omega r)\sin\tilde{\gamma}_\nu \right],
\label{pw70}
\ee
where $J_n$, and $Y_n$ are the Bessel and  Neumann function of $n$-th order, respectively,
and the phases $\tilde{\gamma}_\nu(\omega)$ are to be determined from the
boundary condition on the circle $|\vec{r}|=r_0$.

The asymptotic behaviour of the solution \rref{pw7} at $r\omega\gg
|\nu|+1$ is
\[
\theta_{\omega,\nu}(r)\propto \sqrt{\frac{2}{\pi   r\omega}}
\cos\left[r\omega+\gamma_\nu(\omega)-\frac{\pi |\nu|}{2}-\frac{\pi}{4}\right],
\]
where the scattering phase shift is given by
\be
\gamma_\nu(\omega)=\frac{\pi\omega\sgn \nu}{2}+\tilde{\gamma}_\nu(\omega).
\label{pw8}
\ee
The first term is a phase shift due to the Aharonov-Bohm flux induced
by the vortex current (see. e.g. Ref.~\cite{Sonin97}) which does not depend on the internal state of the vortex.
The second term describes the effect of the vortex core on
the phonon scatterings. We will show in Sec.~\ref{sec:cm5} that this
phase includes a resonance associated with periodic motion of
the vortex (\ref{m2}--\ref{m3}).

The expressions for the relevant cross-sections in terms of the phase
shifts are well known and we do not reproduce them here.

To consider the scattering and the radiation problems we must  use the equations of motion for the fields, with sources that are
formally singular on the circle $|\vec{r}|=r_0$. To avoid this complication, it is
convenient to rewrite the first Maxwell equation \rref{et-popov11a} at $|\vec{r}|=r_0$ as a
matching condition on the solutions inside,
$|\vec{r}|=r_0-0^+$, and outside, $|\vec{r}|=r_0+0^+$, of the effective vortex
core.

The first requirement is that the physical current $\he\vec{E}$ (we
can choose the current component normal to the contour
$|\vec{r}|=r_0$), does not diverge. This condition can be written in
integral form:
\begin{subequations}
\label{pw9}
\be
\int_{\vec{r}_-}^{\vec{r}_+} d\vec{r}\he {\vec E}(\vec{r},t)=0;
\quad
\vec{r}_\pm = (r_0\pm 0^+) \hat{r}
\label{pw9a}
\ee
where $\hat{r}$ is a unit vector in the direction of $\vec{r}$.
Here, and in what follows, the line integration is assumed to be along the straight
line connecting the endpoints $\vec{r}_\pm$.
To obtain the second boundary condition we substitute
\[
\nnabla\he\vec{E}=\frac{\vec{r}}{|\vec{r}|^2}
\left[\nnabla \left(\vec{r}\he \vec{E}\right)-\he \nnabla\left(\vec{r}\cdot \vec{E}\right) \right]
\]
into the Maxwell equation (\ref{et-popov11a}) and integrate the result between the points
$\vec{r}_\pm$ defined in \req{pw9a}, yielding:
\be
\left(\vec{r}\he \vec{E}\right)\Bigg|_{\vec{r}_+}-\left(\vec{r}\he
  \vec{E}\right)\Bigg|_{\vec{r}_-}
= \int_{\vec{r}_-}^{\vec{r}_+} d\vec{r}\left[\he
  \nnabla\left(\vec{r}\cdot \vec{E}\right)-\vec{r}\, \pt B\right].
\label{pw9b}
\ee
\end{subequations}

Next we apply the partial wave expansion to study the
effects of the coupling of the vortex motion to the environment.

\subsubsection{Emission of phonons by an  oscillating vortex, and the resulting decay rate.}
\label{sec:cm4}

Let us consider a vortex with $\sigma=1$ experiencing  weak
oscillations around the origin $|\vec{R}(t)|\ll 1$.
 Linearizing the
right-hand-side of \req{et-popov11b} with respect to small
$\vec{R}(t)$ yields:
\be
\nnabla \cdot \left(\frac{\vec{E}}{B}\right) =
\left(2\pi-\vec{\mathbb{D}}(t)\cdot\nnabla\right)\delta_{r_0}(\vec{r}\,),
\label{et-popov11b-dipole}
\ee
where $\delta_{r_0}$  is the smeared $\delta$-function  defined in
\req{smeareddelta}, and the total dipole moment [compare with \req{et-popov11c}],
\be
\vec{\mathbb{D}}(t)\equiv 2\pi\vec{R}(t)+\frac{\partial}{\partial \f}
\left(\frac{\pi \f^{\, 2}}{\alpha^2}\ln\frac{1}{|\f|}\right),
\label{total-dipole}
\ee
perturbs the surrounding field due to the displacement of the vortex itself (the first term) to which is added the non-analytic deformation of the vortex core (the second term). Clearly, these effects are additive within the framework of a linear
theory.

In analogy with our introductory remarks, (see \req{pw1}), we look for a solution in the form
\be
\frac{\vec{E}}{B}=
\frac{\vec{r}}{|\vec{r}|^2}\Theta\left(|\vec{r}|-r_0\right)
- \vec{\mathbb{D}}(t)\delta_{r_0^-}(\vec{r}\,)
-\he\vec{\nabla}\theta(\vec{r},t)
.
\label{pw1-dipole}
\ee

To find the source term for the density $B$ we use \req{et-popov11d}. We
notice that the magnetic moment contribution due to the vortex
motion \rref{et-popov11f} vanishes to linear order in $\vec{R}(t)$, so that
\be
{\cal M}(\vec{r},t)=\left(\frac{\Theta(r_0-|\vec{r}|)}{\pi r_0^2}
+ \frac{ 2 \vec{R}(t)\cdot\vec{r}}{r_0^2}\delta_{r_0}(\vec{r}\,) \right)\mathfrak{m}^{(c)},
\label{et-popov11e-linearized}
\ee
where the core contribution to the magnetic moment, $\mathfrak{m}^{(c)}$, is given by \req{et-popov11g}. Substituting \reqs{pw1-dipole} and \rref{et-popov11e-linearized} into \req{et-popov11d} yields
\be
\begin{split}
B(\vec{r},t)=&1-
\frac{\Theta(|\vec{r}|-r_0)}{2r^2}-\frac{\Theta(r_0-|\vec{r}|)}{\pi r_0^2}\mu(B_{\infty})\Bigg|_{B_\infty=1}
\\
&- \pt\theta +\frac{\Theta(|\vec{r}|-r_0)\vec{r}}{|\vec{r}|^2}
\he\vec{\nabla}\theta
\\
&
- \delta_{r_0^-}(\vec{r}\,)\frac{2\vec{R}(t)\cdot\vec{r}}{r_0^2}\mathfrak{m}^{(c)}.
\end{split}
\label{pw2-source}
\ee
This equation replaces \reqs{pw2}. The third line in \req{pw2-source} represents the singular source due
to the motion of the vortex. From what follows it will become clear that this magnetic moment term is responsible for the coupling of the vortex motion to the environment.

Using \reqs{pw1-dipole} and \rref{pw2-source} we can identify the
source term in the equation for the electric field.
Neglecting terms smaller by $1/r_0^2$ than unity [such as the last term in \req{pw3a}], we get
\be
\begin{split}
\vec{E}=&\frac{\Theta(|\vec{r}|-r_0)\vec{r}}{r^2}\left[1-\frac{1}{2r^2}\right]
\\
&
- \frac{\Theta(|\vec{r}|-r_0)\vec{r}}{r^2}\pt\theta -\he\nnabla\theta
\\
&
- \vec{\mathbb{D}}(t)\delta_{r_0^-}(\vec{r}\,).
\end{split}
\label{pw3-source}
\ee

\begin{subequations}
\label{pw9source}
Now, we are ready to obtain the modification of boundary conditions \rref{pw9} for the field
$\theta$ on the circle $|\vec{r}|=r_0$. Substituting \req{pw3-source} into
\req{pw9a} yields
\be
\theta(\vec{r}_+)-\theta(\vec{r}_-)=\frac{\vec{r}_+\he \vec{\mathbb{D}}(t)}{|\vec{r}_+|^2}.
\label{pw9source-a}
\ee
Similarly, substituting  \reqs{pw3-source} and \rref{pw2-source} in
\req{pw9b} yields
\be
\left(\vec{r}\nnabla \theta \right)\Bigg|_{\vec{r}_+}-\left(\vec{r}\nnabla \theta\right) \Bigg|_{\vec{r}_-}
=-\frac{\vec{r}_+\he \vec{\mathbb{D}}(t)}{|\vec{r}_+|^2} +\frac{\pt\vec{R}(t)\cdot\vec{r}_+}{\pi r_0^2}\mathfrak{m}^{(c)}.
\label{pw9source-b}
\ee
\end{subequations}

Equations \rref{pw9source} enable us to express the fields outside
and inside the vortex in terms of the total dipole moment
\rref{total-dipole} and the velocity of the vortex, as the solutions
in all space are given by Eqs.~(\ref{pw5}--\ref{pw7}). Moreover, the
symmetry of the sources in \reqs{pw9source} dictate that only terms
with $\nu=\pm 1$ are excited (dipole approximation). Substituting
\req{pw5} into \req{pw9source-a} we obtain the matching condition for
the radial functions
\begin{subequations}
\label{pw9source-radial}
\be
\theta_{\pm 1}(r_0+0^+)-\theta_{\pm
  1}(r_0-0^+)=\pm \frac{{i\mathbb{D}}_{\mp}(\omega)}{2 r_0},
\label{pw9source-a-radial}
\ee
where the complex dipole moment is given by
\be
{\mathbb{D}}_{\pm}(\omega)= {\mathbb{D}}_{x}(\omega) \pm i
{\mathbb{D}}_{y}(\omega);
\ \left[{\mathbb{D}}_{+}(\omega)\right]^*={\mathbb{D}}_{-}(-\omega),
\label{pw9source-b-radial}
\ee
and $ \vec{\mathbb{D}}(t)=\sum_\omega e^{-i\omega t}\vec{\mathbb{D}}(\omega)$.
In a similar fashion \req{pw9source-b} yields
\be
\frac{d}{dr}\theta_{\pm 1}(r_0+0)-\frac{d}{dr}\theta_{\pm
  1}(r_0-0^+)=\mp \frac{{i\mathbb{D}}_{\mp}(\omega)}{2 r_0^2}
 -\frac{i\omega{R}_{\mp}(\omega)}{2\pi r_0^2}\mathfrak{m}^{(c)},
\label{pw9source-c-radial}
\ee
with
\be
R_{\pm}(\omega)= R_{x}(\omega) \pm i
R_{y}(\omega);
\ \left[R_{+}(\omega)\right]^*={R}_{-}(-\omega),
\label{pw9source-d-radial}
\ee
\end{subequations}
and $ \vec{R}(t)=\sum_\omega e^{-i\omega t}\vec{R}(\omega)$.

We can now complete our solution of the radiation problem. We use solutions \rref{pw6b}
and \rref{pw7} for $m\pm 1$, and the asymptotics of the Hankel function
\[
\frac{i\pi}{2}H^{(1)}_{1+\eta}(x)=\frac{1}{x}\left(\frac{2}{x}\right)^\eta
+ \frac{i\pi}{4}x\left(\frac{x}{2}\right)^\eta
\]
for $x,\eta \ll 1$. (The second term in this expansion is imaginary and
therefore it should be kept although it is small in comparison with first large - but real -
term).
As $\omega r_0 \ll 1$, \reqs{pw9source} can be solved by iterations.
Neglecting first the field inside the vortex core $r<r_0$, we obtain
from \req{pw9source-c-radial} the field in the exterior domain:
\be
\theta_{\pm 1}(r>r_0)=\frac{\pi \omega}{4}\left(\frac{\omega
    r_0}{2}\right)^{\pm\omega} H^{(1)}_{1\mp \omega}(\omega r)
\left[\mp {\mathbb{D}}_{\mp}(\omega)
 -\frac{\omega}{\pi}{R}_{\mp}(\omega)\mathfrak{m}^{(c)}
\right],
\label{theta-outside-radiation}
\ee
where we assumed $\omega >0$.
Next, we use \req{pw9source-b-radial}, keeping only terms with real
coefficients (as these are the only terms which can lead to decay of the
vortex motion). Then, \req{pw6b} yields
\be
\theta_{\pm 1}(r<r_0)=\frac{\pi \omega^2r}{8}\left(\frac{\omega
    r_0}{2}\right)^{\pm 2\omega}
\left[\mp {\mathbb{D}}_{\mp}(\omega)
 -\frac{\omega}{\pi}{R}_{\mp}(\omega)\mathfrak{m}^{(c)}
\right].
\label{theta-inside-radiation}
\ee
Now we can find the effective fields which enter into the equation of motion for the vortex
\rref{et-vortex-motion}. Using \reqs{pw3}, we obtain
\begin{subequations}
\label{fields-inside}
\be
E_{\pm}\equiv E_x\pm i E_y=
\frac{i\pi\omega^2}{4}\left(\frac{\omega
    r_0}{2}\right)^{\mp 2\omega}
\left[-{\mathbb{D}}_{\pm}(\omega)
 \pm \frac{\omega}{\pi}{R}_{\pm}(\omega)\mathfrak{m}^{(c)}
\right]
\label{fields-inside-a}
\ee
for the electric field, and
\be
 \nabla_{\pm} B = \frac{i\pi \omega^3}{4}\left(\frac{\omega
    r_0}{2}\right)^{\mp 2\omega}
\left[\pm {\mathbb{D}}_{\pm}(\omega)
 -\frac{\omega}{\pi}{R}_{\pm}(\omega)\mathfrak{m}^{(c)}
\right]; \quad \nabla_{\pm}\equiv\nabla_x\mp i\nabla_y,
\label{fields-inside-b}
\ee
for the density.
\end{subequations}

Equations \rref{fields-inside} can be further simplified with the help
of the equations of motion of the vortex \rref{et-vortex-motion} and the
expression of for total dipole moment \rref{total-dipole}.
Differentiating \req{total-dipole} over time, expressing $\pt \vec{R}$
in terms of the force \rref{force2}, and using \req{et-vortex-motion} we
find
\be
\pt \vec{\mathbb{D}}=\he \vec{E} + \frac{\he\nnabla{ B }}{2\pi}\mathfrak{m}^{(c)}.
\label{dipole-moment-equation}
\ee
Substituting \req{dipole-moment-equation} and $\pt
\vec{R}=\he\vec{E}-\he\vec{f}$, keeping only leading in $\omega\ll
1$ terms, and allowing for negative $\omega$ we obtain simple expressions
\be
\begin{split}
&E_{\pm}=
\frac{-i\pi{f}_{\pm}(\omega)\omega|\omega|}{4\pi}\left(\frac{|\omega|
    r_0}{2}\right)^{\mp 2\omega}\mathfrak{m}^{(c)}; \quad f_{\pm}=f_x\pm i f_y;
\\
& \nabla_{\pm} B = \frac{\mp i\pi{f}_{\pm} \omega^2|\omega|}{4\pi}
\left(\frac{|\omega|
    r_0}{2}\right)^{\mp 2\omega}\mathfrak{m}^{(c)}.
\end{split}
\label{fields-inside-simple}
\ee

\begin{figure}[h]
\begin{center}
\includegraphics[width=0.6 \columnwidth]{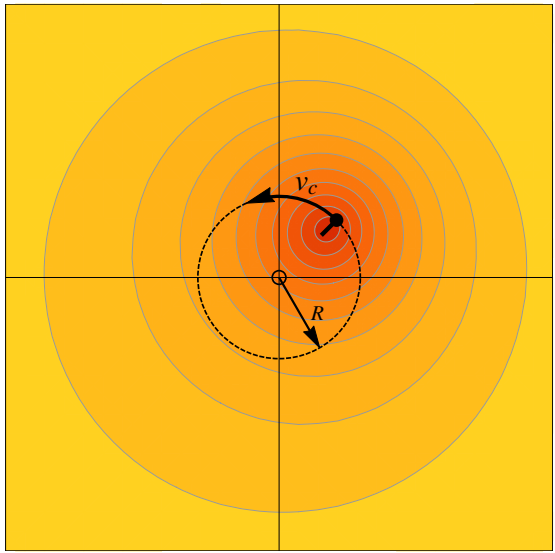}
\caption{Rotating vortex (black dot) and its dipole moment (cut). At
  the distances larger than the healing length the density restores
  its form for a motionless vortex at $R=0$.}
\label{fig-d}
\end{center}
\end{figure}

It is worth emphasizing that the dipole moment $\mathbb{D}$ has
disappeared from the equations for the field. This is because
according to equation of motion of the decoupled vortex \rref{m1},
the dipole moment due to the motion of the vortex itself is
compensated by the dipole moment of the cut, as illustrated in Fig.~\ref{fig-d}.

Finally, we are ready to solve the equation of motion
\rref{et-vortex-motion} taking into account the self-consistent field
created by the vortex motion.
Looking for solution in a form similar to \req{m2}
\be
f_+\equiv f_x+if_y=Fe^{{i\Omega^* t}}; \quad f_-\equiv f_x-if_y=F^*e^{-{i\Omega t}},
\label{m2-modified}
\ee
we obtain from \req{et-vortex-motion}
\be
\left[\frac{\Omega}{\Omega_c(|F|)}-1\right]f_-=\frac{\nabla_-B}{\pi}\mathfrak{m}^{(c)},
\label{almost-spectrum}
\ee
where the frequency of the oscillation of the vortex decoupled from
the phonons, $\Omega_c(|F|)$, is given by \req{m3}.
Substituting the second of \reqs{fields-inside-simple}
into \req{almost-spectrum}, we obtain
\be
\Omega(|F|)=\Omega_c(|F|)+\frac{i}{2\tau[\Omega(|F|)]};
\quad
\frac{1}{\tau(\omega)}= \frac{\omega^4}{2\pi}\left[\mathfrak{m}^{(c)}\right]^2,
\label{spectrum}
\ee
where we neglect the factor $\left(\omega
    r_0/{2}\right)^{ 2\omega}\simeq 1$ in the
expression for the decay rate $1/\tau$.

The frequency dependence of the relaxation rate, $\tau_c^{-1}\propto {\omega^4}$,
deserves some discussion. Were the radiation simply that of
a dipole $\propto R$ in two-dimensional electrodynamics, one would obtain
radiation power $\simeq R^2|\omega^3|$. As the kinetic energy (in our
units) is proportional to $\omega^2R^2/\Omega_c$, it would imply $1/\tau
\propto \Omega_c^2$. However, as we mentioned above  (see also
Fig.~\ref{fig-d}), the effective dipole moment vanishes and the only
reason for the remaining coupling is the time dependence due to the motion of the magnetic
moment $\mathfrak{m}^{(c)}$. Now, as is well known from classical electrodynamics, the power of magnetic moment radiation is smaller than that of electric dipole radiation by an extra second power of the frequency (in the low frequency limit).
This is the reason for the $\Omega_c^4$ dependence of the relaxation rate.

A last point we should discuss is the value of the core magnetic
moment $\mathfrak{m}^{(c)}$. At $r_0\gg 1$, the energy of the vortex
currents that accumulated up to a distance $r_0$ is $\Delta
E\approx \pi B \ln(r_0\sqrt{B})$. This would imply that the value of the
core magnetic moment is scale dependent
\be
\mathfrak{m}^{(c)} =\pi\ln r_0;
\label{excuses1}
\ee
Getting rid of the scale dependence requires consideration of the
current near the vortex with higher accuracy [perturbative
treatment of the last term in \req{pw3a} and the right-hand-sided in
\req{pw4a}]. Such a treatment leads to a replacement of $\ln r_0 \to \ln(1/\Omega_c) \simeq
\ln \ln(1/|F|)$ in this
estimate. We do not believe that this double logarithmic dependence is
relevant or observable and pursue this issue no further.

\subsubsection{Elastic scattering of phonons by an oscillating vortex.}
\label{sec:cm5}

To study the scattering problem one should consider the field outside the
effective vortex core to be in the form of a cylindrical wave, see \req{pw70}. As we already
saw, for the weakly oscillating vortex only the dipole mode, $\nu=\pm 1$,
contributes, and in what follows we shall focus our attention on the $\nu = 1$ mode:
\be
\theta_{1,\omega}(r)= A_{1,\omega}\left[J_{1-\omega}(\omega r)\cos\tilde{\gamma}_1 -Y_{1-\omega}(\omega r)\sin\tilde{\gamma}_1 \right],
\label{external-wave}
\ee
where the wave amplitude  $A_{1,\omega}$ is assumed to be sufficiently small so
that all the equations can be linearized with respect to this amplitude.

Let us neglect the decay of the free vortex motion due to the
radiation and study the effect of the additional wave field
\rref{external-wave} on the vortex motion:
\be
\begin{pmatrix}f_-\\
R_-\\
\mathbb{D}_-
\end{pmatrix}
=\begin{pmatrix}
F^*\\
R^*\\
\mathbb{D}^*
\end{pmatrix}e^{-{i\Omega_c t}}
+
\begin{pmatrix}\delta f_-
\\
\delta R_-
\\
\delta \mathbb{D}_-
\end{pmatrix}e^{-i\omega t}.
\label{m2-modified-force}
\ee
where all the coefficients $\delta\cdot$ are linear in
$A_{1,\omega}$.

Linearizing \req{et-vortex-motion} with respect to the small perturbation
$\delta\cdot$ yields
\be
\left[\frac{\omega}{\Omega_s(|F|)}-1\right]\delta f_-=\frac{\nabla_-\delta B}{\pi}\mathfrak{m}^{(c)}
\label{vortex-motion-scattering}
\ee
where $\nabla_-$ is defined in \req{fields-inside-b}. The
eigen-frequency for this small oscillation $\Omega_s(|F|)$ is
different from the effective cyclotron frequency $\Omega_c(|F|)$ due to
the logarithmic dependence of the frequency on the period of the oscillations:
\be
\frac{1}{\Omega_s(|F|)}=\frac{1}{\Omega_c(|F|)}-\frac{1}{2}.
\label{frequency-scattering}
\ee

In analogy with our solution of the radiation problem, we utilize \reqs{pw9source-radial}
and the asymptotics of the Bessel and  Neumann functions
\[
Y^{(1)}_{1+\eta}(x)=\frac{x}{2}\left(\frac{x}{2}\right)^\eta;\quad
Y^{(1)}_{1+\eta}(x)=-\frac{2}{\pi x}\left(\frac{2}{x}\right)^\eta;
\]
for $x,\eta \ll 1$.

Instead of \req{theta-outside-radiation}, we obtain
\be
A_{1,\omega}\sin\tilde{\gamma}_1=\frac{i\pi\omega}{4}\left(\frac{2}{\omega
    r_0}\right)^{\omega}
\left[ \delta{\mathbb{D}}_{-}
 +\frac{\omega}{\pi}\delta{R}_{-}\mathfrak{m}^{(c)}
\right],
\label{theta-outside-scattering}
\ee
and then \req{theta-inside-radiation} gives
\be
\theta_{1}(r<r_0)
=\frac{r\omega}{2}A_{1,\omega}\cos\tilde{\gamma}_1\left(\frac{2}{\omega
    r_0}\right)^{\omega}.
\label{theta-inside-scatterring}
\ee

Equation \rref{theta-inside-scatterring} immediately produces,
\be
\delta E_-=-i\omega A_{1,\omega}\cos\tilde{\gamma}_1\left(\frac{2}{\omega     r_0}\right)^{\omega}; \quad
\nabla_{-} \delta B = i\omega^2 A_{1,\omega}\cos\tilde{\gamma}_1\left(\frac{2}{\omega
     r_0}\right)^{\omega},
\label{fields-inside-scatterring}
\ee
Repeating the same arguments as those leading to
\reqs{fields-inside-simple}, we find
\be
A_{1,\omega}\sin\tilde{\gamma}_1=\frac{i\pi\omega\delta f_-}{4\pi}\left(\frac{2}{\omega
    r_0}\right)^{\omega}
\mathfrak{m}^{(c)}.
\label{theta-outside-scattering-simplified}
\ee

Substituting \reqs{fields-inside-scatterring} and \rref{theta-outside-scattering-simplified}
into the linearized vortex equation of motion
\rref{vortex-motion-scattering} we obtain
\[
\left[\frac{\omega}{\Omega_s(|F|)}-1\right]\delta f_-=-\frac{\delta
  f_-}{2 \omega\tau(\omega)}\cot \tilde{\gamma}_1(\omega),
\]
where the decay rate $1/\tau(\omega)$ is given by \req{spectrum}.

\begin{figure}[h]
\begin{center}
\includegraphics[width=0.7\columnwidth]{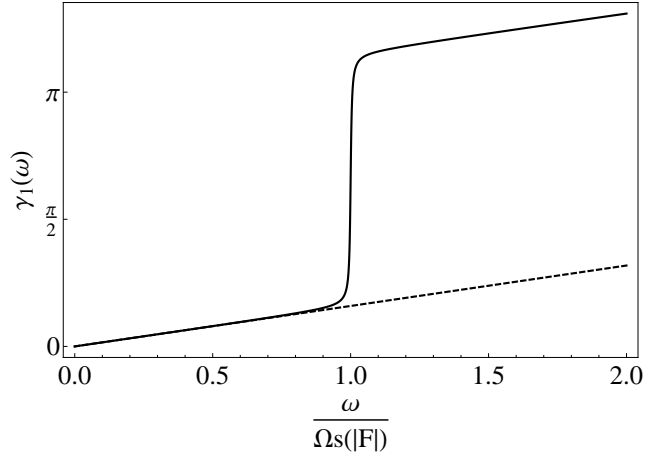}
\caption{The elastic scattering phase in $\nu=1$ channel as the function
  of frequency of the incoming wave. The dashed line is the
  Aharonov-Bohm phase [first term in \req{scattering-phase}]. The
  narrow jump corresponds to the resonance scattering by the
  oscillation of the vortex.}
\label{fig:scatteringphase}
\end{center}
\end{figure}

Finally, using \req{pw8}, we obtain the complete expression for the
scatterring phase:
\be
\gamma_1(\omega)=\frac{\pi\omega
}{2}+\arctan\left(\frac{\Omega_s(|F|)}{2\left[\Omega_s(|F|)-\omega\right]\omega\tau(\omega)}\right)
\label{scattering-phase}
\ee
where we have chosen $0<\arctan(x)<\pi$. The overall phase dependence is
shown in Fig.~\ref{fig:scatteringphase}.

The first term in the right hand side of the above equation comes from scattering on Aharonov-Bohm flux created by the vortex. This has been discussed extensively in the literature, see e.g. Ref.~\cite{Sonin97}. The second
term is due to the excitation of the vortex into circular motion by the
incident waves and, then, re-emission of this excitation. Far from
resonance this contribution is negligible, however, when
$\omega\to\Omega_s(|F|)$, it produces unitary scattering in the $\nu=1$
channel. This resonance is narrow, and, therefore, its contribution is
most likely irrelevant for the scattering of phonons with a
broad quasiequilibrium distribution. However, when the system is out of equilibrium, the
scattering of phonons on the vortices produces a distribution
function with a singularity at the frequency of the vortex
motion. These non-equilibrium phonons are strongly scattered on other vortices, providing a kinetic mechanism for vortex-vortex interaction, as the relaxation rate of the phonon energy is
extremely long. We hope to come back to a detailed study of this mechanism in the nearest future.

\section{Semiclassical quantization of the vortex motion.}
\label{sec:qm}

In the previous section we established that the cyclotron motion of
the excited vortex is a long living excitation in the classical sense, see
\req{spectrum}, i.e. the classical motion is almost single-periodic. In
quantum mechanics a single-periodic motion corresponds to discrete
excitation levels, and the classical decay rate becomes the
characteristic broadening of those levels.

To calculate the position of the discrete levels ${\cal E}^V_j$, it is convenient to use the quantum-to-classical correspondence principle
(differential form of the Bohr-Sommerfeld quantization condition)
\be
{\cal E}^V_{j+1}-{\cal E}^V_j=\hbar\Omega_c({\cal E}^V_j);
\label{qm1}
\ee
where $\Omega_c({\cal E}^V_j)$ is the frequency of the periodic motion at
energy  ${\cal E}^V_j$. For harmonic oscillators, the frequency
$\Omega_c({\cal E}^V_j)$ is independent of energy and \req{qm1} gives the exact
spectrum. In our case, the frequency of the periodic motion $\Omega_c$
from \req{m3} is amplitude dependent and we need to express $|\f|$ in
terms of the energy of the system.

To accomplish this  task,
we consider the effective action \rref{action2-et} --
\rref{Vortex-core-action} for one vortex. Fixing $\vec{E}=0$, $B=1$
inside the vortex, and restoring the physical units as described after \reqs{hydrodynamic},
we find the effective action of the vortex:
\be
{\cal S}_V=\int dt {\cal L}_V,
\ee
with the Lagrangian
\be
{\cal L}_V=\frac{(n_0m\xi^2)\pi }{\alpha^2}(\pt\vec{R})^2 \ln
\frac{\hbar}{m\xi|\pt\vec {R}|}-
2\pi  \vec{a}(\vec{R}(t))\cdot \partial_t \vec{R}(t),
\label{qm2}
\ee
where $\nnabla\times \vec{a}=\hbar n_0$.
Thus the energy of the vortex, ${\cal E}^V$, is given by
\be
{\cal E}^V=\partial_t \vec{R}\cdot \frac{\partial {\cal L}_V }{\partial (\partial_t
  \vec{R})}-{\cal L}_V=
\frac{(n_0m\xi^2)\pi }{\alpha^2}(\pt\vec{R})^2 \left[\ln
\frac{\hbar}{m\xi|\pt\vec {R}|}-1\right];
\label{qm3}
\ee

The cyclotron frequency \rref{m3} in the original units takes the form,
\be
\Omega_c =\frac{\hbar\alpha^2}
{m\xi^2\left[\ln \frac{\hbar}{m\xi|\pt\vec {R}|}-\frac{1}{2}\right]};
\label{qm4}
\ee
Thus for circular motion, $(\pt\vec {R})^2=\Omega(E)^2R^2(E)$,
we obtain, with logarithmic accuracy,
\be
R^2({\cal E}^V_{j+1})-R^2({\cal E}^V_{j})=\frac{1}{n_0\pi};
\label{qm5}
\ee
with the solution
\be
R^2({\cal E}^V_{j})=\frac{j+1/2}{n_0\pi};
\label{qm6}
\ee
(The additive term $1/2$ is exact for harmonic oscillator and
should not be that different for our problem as the non-linearity is
only logarithmical). We see that the amplitude of the quantum
mechanical oscillation of the vortex position for the ground state of
the vortex, $j=0$, is precisely the interparticle distance.

Finally, substituting \req{qm6} into \req{qm3} we obtain with
logarithmic accuracy the discrete levels
\be
{\cal E}^V_j=\frac{\hbar^2\alpha^2}
{m\xi^2}\left[\frac{2j+1}{\ln \frac{\pi n_0\xi^2}{\alpha^4 (j+1/2)}}\right].
\label{qm7}
\ee
(The energy of zero-point motion $j=0$ can not be separated form the much
larger mean-field core energy). For typical Bose condensates made of Rb atoms the healing length is of order of a micron and the effective two dimensional density is typically $5\cdot 10^{9} \mbox{cm}^{-2}$, see e.g. Ref.~\cite{Coddington03}). Thus the excitation energy of the vortex is of order of 40Hz.

Equation \rref{qm7} is the main result of this section. It gives a
complete description of the quantized energy levels of the vortex
motion. We emphasize that the condition of weak
interaction, ${n_0\xi^2}=1/\lambda \gg 1$, must be fulfilled in order for the
energy levels \rref{qm7} to be within the applicability range of the theory.
In the limit of strong interaction all these levels are not
relevant for the low energy spectrum.

Finally, we address the question of inter-level transitions due to
spontaneous phonon-emission. The rate of this transition can be extracted directly from the
classical relaxation rate \rref{spectrum} by the requirement
that the energy loss calculated quantum-mechanically coincide with the
classical calculation. Restoring the units defined after
\reqs{hydrodynamic},
we find
\be
\frac{1}{\tau_{(j+1)\to j}}=\left(\frac{{\cal E}^V_j}{2\pi\hbar}\right)
\left(\frac{m\xi^2\left({\cal E}^V_{j+1}-{\cal E}^V_{j}\right)}{\hbar^2}\right)^3
\left[\mathfrak{m}^{(c)}\right]^2
\label{qm8}
\ee
The absorption rate and rate of the induced emission can be found from
\req{qm8} using the usual Einstein relation and the equation for the
kinetics of level occupation can be obtained.

\section{Conclusions and outlook.}
\label{sec:conclusions}

In this paper we studied the deformation of the internal structure of the vortex in
two-dimensional superfluid due to a force, $\f$, acting on the vortex.
Contrary to the conventional expectation that such deformation is
small and analytic in $\f$, we showed in Sec.~\ref{sec:solution} that
it is, in fact, non-analytic, and results in an
anomalous dipole moment of the vortex $\propto \f\ln(1/|\f|)$.
We incorporated the effects of this dipole moment into the effective
theory of superfluid defined on distances larger than the healing length,
see Sec.~\ref{sec:eft}. Armed with this effective theory,
we investigated the
dynamics of the vortex and its coupling to the phonons, see
Sec.~\ref{sec:cm}. We found that the oscillatory
motion of the vortex is characterized by low
frequency, see \req{m3}, which depends logarithmically on the
amplitude of the oscillations and increases alongside it. We also found that the oscillations of the vortex have a long
classical lifetime, see \req{spectrum}, and thus can be quantized, see
Sec.~\ref{sec:qm}.

The lower bound on the oscillation frequency of the mode is determined
by the ratio of the amplitude of the quantum motion of the vortex (which is of the
order of the inter-particle distance) to healing length. In the strong coupling region,
such as  in $^4\mbox{He}$ superfluid, the healing length and the inter-particle distance are of
the same order and the excitations considered in this paper are irrelevant since they are of
high-energy and short lifetime. On the other hand, for the weakly interacting system (which
may be  realized, {\em e.g.}, using cold atoms) the healing length is much larger than the inter-particle distance, and the vortex excitations are well within the phonon spectrum. In this limit, the energy levels of the excited vortex may affect the phonon kinetics, because the energy dependence of the elastic phonon scattering on the vortices acquires resonance features, see Sec.~\ref{sec:cm5}.

As a parting statement, we enumerate some possible further developments
on effects of the non-analytic core deformation.

\begin{enumerate}

\item {\em Internal structure of vortex-antivortex pairs.}
In this paper we considered only vortices separated by distances
much larger than the healing length, and therefore the deformation of their cores
could be considered independently.  However, when vortices come closer together the situation becomes more complicated. To illustrate, if one considers an vortex-antivortex pair (separated by a distance $R$) at rest, each of them will  experience a Magnus force due to the superfluid flow created by the other vortex.  As a result, each vortex will produce a cut directed at it pair. As the intervortex distance goes down, the Magnus force grows and so does the size of the cuts. Our
preliminary study \cite{We1} indicates that at $R\lesssim 10$ the
solution of the two cuts becomes less favorable than a
``string'' solution (a line of zero density connecting the vortex and antivortex).
 The quantum tunnelling between those two stable solutions (``quantum string breaking'') as
well as further annihilation of the vortex-antivortex pair should be
important for the study of vortex kinetics and of the effects of
vortex-antivortex dipoles on the phonon spectra.

\item {\em Strong non-linearity for vortices driven by monochromatic
  radiation.} If we neglect quantum effects, the
frequency of oscillation $\Omega_c$ is a an increasing function of the oscillation amplitude $F$, see \req{m3}. Consider, now, the application of monochromatic
radiation with amplitude $A$ and the frequency $\omega$, on the vortex.
It is known from the classical mechanics
(Duffing oscillator) that such a system contains solutions corresponding to small oscillations $F\propto A$, but also solutions determined by the condition $\Omega(F)\simeq\omega$, see e.g. \cite{Dykman12}. This latter solution turns
out to be stable. Thus, in the absence of quantum effects arbitrarily small forces lead to a finite amplitude of the
vortex oscillations. Quantum mechanics (see Sec.~\ref{sec:qm}) puts a
lower bound on frequency $\omega>\Omega_{1}-\Omega_{0}$, and on the
amplitude of the force. If those restriction are fulfilled the
non-linearity of the vortex dynamics still takes place, and reveals itself in strongly
non-equlibrium peaked distribution function over energy levels of the vortex.
The kinetics of such processes deserves further study.

\item {\em Generalization to three dimensional systems and higher spin
  bosons.} Our analysis was deliberately limited to the case of
spinless bosons in two spatial dimensions where the description of the vortex
singularities was somewhat trivial [vortex line in 2+1 dimensional
space]. The obvious generalizations of Popov's formalism are to space of higher
dimension [vortex is a surface in 3+1 dimensional
space] and to bosons with higher spin [vortex is characterized not
only by its position but also by the direction in the spin
sector\cite{SpinorSuperfluidReview}]. Even though the precise mathematical tools should be
developed we believe that the requirement of the non-anlytic
reconstruction of the vortex core is robust and the slow
oscillatory motion of the vortex should reveal itself for such systems as well.
\end{enumerate}

\section*{Acknowledgement}

We are grateful to 
V. Cheianov for valuable discussions on the initial stage of this work and to
R. Fattal and D. Klein for help and advice on the numerical work in this paper.
We also thank A. Abanov, Y. Galperin,  N. Prokofiev, and L. Radzihovsky for
reading the manuscript and remarks, and to L. Glazman and  A. Kitaev
for a discussion of the results. 
 This research has been supported by the United
States-Israel Binational Science Foundation (BSF) grant
No.~2012134 (O.A. and A.K.) and Simons foundation
(O.A. and I.A.).

\appendix

\section{The numerical procedure}
\label{appendixA}

In Sec. \ref{sec:numer-analys-popovs} we gave an overview of our solution of Popov's equations. This consisted of solving the fictitious-time dependent equations, \eqref{eq:GPE-time} and \eqref{eq:sGPE-1}. In what follows, we give a step-by-step description of the algorithm.

The discrete form of the equations that we solve are:
\begin{align}
  D_x e_x^{k+1,i-1/2,j} + D_y e_y^{k+1,i,j-1/2} & = -\frac{2\pi}{\Delta^2} \delta_{i,{R_v}_x}\delta_{j,{R_v}_y} \label{eq:sP-discrete1-app}\\
D_t \rho^{k,i,j} &= \mathcal{H}^{k,i,j}\rho^{k+1,i,j}.\label{eq:sP-discrete-2-app}
\end{align}
Here, $f^{t,x,y}$ denotes the value of $f$ at time $t$ and positions $x,y$ on the lattice. $\delta_{i,j}$ is the Kronecker delta. The various difference operators are defined by
\begin{subequations}
\begin{align}
  \label{eq:diff-ops-app}
  D_xf^{k,i,j} & = \Delta^{-1}(f^{k,i+1,j}-f^{k,i,j}),\\
  D_yf^{k,i,j} & = \Delta^{-1}(f^{k,i,j+1}-f^{k,i,j}),\\
  D_tf^{k,i,j} & = \Delta_t^{-1}(f^{k+1,i,j}-f^{k+1,i,j}),
\end{align}
where $\Delta,\Delta_t$ are, respectively, the lattice constant and time step. The electric field density $\vec{e} = \vec{E} / B
$ is given by
\begin{align}
  \label{eq:velocities-def-app}
  e_x^{k,i+1/2,j} & = \frac{D_x a_0^{k,i,j}}{(\rho^{k,i+1/2,j})^2},\\
  e_y^{k,i,j+1/2} & = \frac{D_x a_0^{k,i,j}}{(\rho^{k,i,j+1/2})^2},
\end{align}
and $\mathcal{H}$ is given by:
\begin{equation}
  \label{eq:H-discrete-app}
  \mathcal{H}^{k,i,j}=\frac{1}{2}\Delta_L - \frac{1}{2}[(e_x^{k,i,j})^2+(e_y^{k,i,j})^2] + 1 - (\rho^{k,i,j})^2
\end{equation}
with $\Delta_L$ a discrete Laplace operator,
\begin{equation}
  \label{eq:Laplace-discrete-app}
  \Delta_L f^{k,i,j}= \Delta^{-2} (f^{k,i+1,j} + f^{k,i-1,j} + f^{k,i,j-1} + f^{k,i,j-1}-4f^{k,i,j}).
\end{equation}
\end{subequations}
In these definitions, difference operators in space shift their operand fields by a half-lattice-constant, creating a staggered grid. Applying another difference operator returns the operands to the integer lattice. An object defined on one lattice is interpolated to the shifted lattice by averaging.

Each of Eqs.~\eqref{eq:sP-discrete1-app} and \eqref{eq:sP-discrete-2-app} is a linear system of equations for one of the fields. Solving Eq. \eqref{eq:sP-discrete1-app} yields the unknown field $a_0^{k+1,i,j}$, and solving Eq. \eqref{eq:sP-discrete-2-app} yields $\rho^{k+1,i,j}$. Each of these systems depends on the data at the previous timestep. For completeness, we write down the full equations. Eq. \eqref{eq:sP-discrete1-app} reads:
\begin{align}
  \frac{1}{\Delta}\left[ \frac{\frac{1}{\Delta}\left(a_0^{k+1,i+1,j}-a_0^{k+1,i,j}\right)}{n^{k,i+1/2,j}} - \frac{\frac{1}{\Delta}\left(a_0^{k+1,i,j}-a_0^{k+1,i-1,j}\right)}{n^{k,i-1/2,j}}\right]+ \left[i\leftrightarrow j\right] \nonumber \\ = -\frac{2\pi}{\Delta^2}\delta_{i,{R_v}_x}\delta_{j,{R_v}_y} \label{eq:sP-discrete1-app-full}
\end{align}
where the $i\leftrightarrow j$ symbol implies that we switch the \emph{function} of the $i,j$ - i.e. we difference in $j$ and not in $i$. The averaged density is given by:
\begin{subequations}
\begin{align}
  \label{eq:n-avg-x}
  n^{k,i+1/2,j} = \left(\frac{\rho^{k,i+1,j}+\rho^{k,i,j}}{2}\right)^2 \\
n^{k,i,j+1/2} = \left(\frac{\rho^{k,i,j+1}+\rho^{k,i,j}}{2}\right)^2\label{eq:n-avg-y}
\end{align}
\end{subequations}

Equation \eqref{eq:sP-discrete-2-app} reads:
\begin{align}
  \label{eq:sP-discrete2-app-full}
&\frac{1}{2\Delta^2}\left( \rho^{k+1,i+1,j} + \rho^{k+1,i-1,j} + \rho^{k+1,i,j+1} + \rho^{k+1,i,j-1} - 4\rho^{k+1,i,j}\right) \nonumber \\
&+ \left[1-(\rho^{k,i,j})^2\right]\rho^{k+1,i,j} - \frac{1}{2}[(e_x^2)^{k,i,j}+(e_y^2)^{k,i,j}]\rho^{k+1,i,j} \nonumber \\
&\qquad = \frac{1}{\Delta_t}\left( \rho^{k+1,i,j} - \rho^{k,i,j} \right)
\end{align}
where
\begin{subequations}
  \begin{align}
  \label{eq:velocities-def-app-2}
  e_x^{k,i+1/2,j} & = \frac{\frac{1}{\Delta}(a_0^{k+1,i+1,j}-a_0^{k+1,i+1,j})}{n^{k+1,i+1/2,j}},\\
  e_y^{k,i,j+1/2} & = \frac{\frac{1}{\Delta}(a_0^{k+1,i,j+1}-a_0^{k+1,i,j})}{n^{k+1,i,j+1/2}},
\end{align}
and
\begin{align}
  \label{eq:v-avg}
  (e_x^2)^{k,i,j} & = \frac{(e_x^{k,i+1/2,j})^2+(e_x^{k,i-1/2,j})^2}{2},\\
  (e_y^2)^{k,i,j} & = \frac{(e_y^{k,i,j+1/2})^2+(e_y^{k,i,j-1/2})^2}{2}.
\end{align}
\end{subequations}
Note that we have used $a_0^{k+1,\ldots}$ in Eq. \eqref{eq:velocities-def-app-2}. The reason for this (as we detail in a moment), is that in practice we solve for $\rho^{k+1,i,j}$ before solving for $a_0^{k+1,i,j}$ and so we can already use the data at time $k+1$ when solving for $a_0$ in Eqs. \eqref{eq:sP-discrete1-app-full}

We initiate our simulation by choosing some initial conditions at $k=0$. Usually we chose $\rho^{0,i,j} = 1$, or alternatively started from some precalculated solution (see later on for some details), and solved eq. \eqref{eq:sP-discrete1-app-full} once to get $a_0^{0,i,j}$. Then we repeat the following procedure step by step:
\begin{enumerate}
\item Solve eq. \eqref{eq:sP-discrete2-app-full} to get $\rho^{k+1,i,j}$;
\item Using the just found value of $\rho$, solve eq. \eqref{eq:sP-discrete1-app-full} to get $a_0^{k+1,i,j}$;
\item $k\to k+1$;
\item \label{item:1} Repeat until $\left|\frac{\mathcal{S}_P^{k+1} - \mathcal{S}_P^{k}}{\mathcal{S}_P^{k+1}}\right| < \varepsilon $;
\end{enumerate}
In the last line, $\mathcal{S}_P^{k}$ is the numerically computed free-energy (\ref{action2}) and $\varepsilon \ll 1 $ is some stopping parameter.

As mentioned in the body of the text, we used MATLAB\texttrademark 's stabilized-biconjugate-gradients method for the actual solving of the linear systems. This choice was one of convenience and we don't believe that it is necessarily better than other accepted sparse system solvers such as SOR. We also used preconditioning to enhance convergence. We used MATLAB\texttrademark 's incomplete LU decomposition to generate preconditioning matrices for the operators in eqs. \eqref{eq:Laplace-discrete-app}+\eqref{eq:velocities-def-app-2}. We omit further details of these minor points for brevity's sake.

We found out, by trial and error, that we could get much better convergence and stability if instead of starting from the high energy initial condition $\rho = 1$, we started from a partially relaxed system. Our usual choice was to solve eqs. \eqref{eq:sP-discrete1-app-full} and \eqref{eq:sP-discrete2-app-full} as described, however with the following change: we replaced $e_x^2,e_y^2$ in Eq. \eqref{eq:sP-discrete2-app-full} with the analytic solution for an unperturbed vortex, i.e. $e_x^2 +e_y^2 = 1/r^2$, with $r$ the radial distance. This proved so helpful that we practically hardwired it into our code, so most (possibly all) of the data in this work was obtained by this procedure.

\begin{figure}[h]
  \centering
\includegraphics[width=0.7\columnwidth]{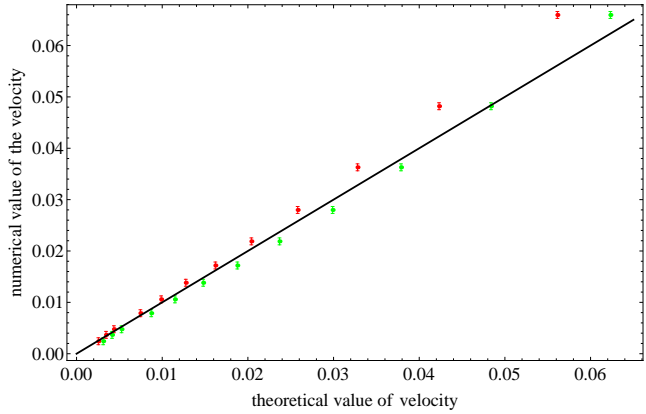}
  \caption{The numerical value of the velocity, extracted from the cut's length, versus the actual velocity at the point of the vortex. The green and the red marks are the velocities calculated with and without the logarithmic shift of virtual position of the vortex, respectively. The error bars are due to the finite size of lattice-spacing which limits the accuracy of the measured cut's length (Additional error which comes form the finite size of the system is not included).}
  \label{fig:cut-scaling}
\end{figure}

The central result of these numerical simulations was the appearance and scaling of the cut. This behaviour is shown in Figs.~\ref{fig1}, \ref{fig2} and \ref{fig:cut-scaling}. As a last point, we note that the cutoff point of fig. \ref{fig:cut-scaling} is not accidental. Rather, at higher flows a new phenomenon appears. This is the the vortex-antivortex collapse mentioned in our concluding remarks in the body of the paper.

\begin{thebibliography}{99}
\bibitem{Putterman74} S.J.~Putterman, {\it Superfluid hydrodynamics (North-Holland series in low temperature physics)}, North-Holland, 1974.

\bibitem{Sonin2013} E.B. Sonin, {\it Transverse force on a vortex and vortex mass: Effects of free bulk and vortex-core bound quasiparticles} Phys.  Rev. B, {\bf 87}, 134515 (2013).

\bibitem{Henkel2010} N. Henkel, R. Nath, and T. Pohl, {\it Three-Dimensional Roton Excitations and Supersolid Formation in Rydberg-Excited Bose-Einstein Condensates}, Phys. Rev. Lett.  {\bf 104}, 195302 (2010).

\bibitem{Arrigoni2013} F. Arrigoni, E. Vitali, D.E.Galli, and L. Reatto, {\it Excitation spectrum in two-dimensional superfluid} $^4$He, Low Temp. Phys. {\bf 39}, 793 (2013).

\bibitem{Anderson95}, M.H. Anderson, J.R. Ensher, M.R. Matthews, C.E. Wieman, and E.A. Cornell, {\it Observation of Bose-Einstein condensation in a dilute atomic vapor} Science {\bf 269}, 198 (1995).

\bibitem{Bradley95} C.C. Bradley, C.A. Sackett, J.J. Tollett, R.G. Hulet, {\it Evidence of Bope-Einstein condensation in an atomic gas with attractive interactions},  Phys. Rev. Lett. {\bf 75}, 1687 (1995).

\bibitem{Davis95} K.B. Davis, M.O. Mewes, M.R. Andrews, N.J. Vandruten, D.S. Durfee, D.M. Kurn and W. Ketterle,{\it Bose-Einstein condensation in gas of sodium atoms}, Phys. Rev. Lett. {\bf 75}, 3969 (1995).

\bibitem{vinen02} W.F. Vinen, and J.J. Niemela, {\it Quantum turbulence}, J. Low. Temp. Phys. {\bf 128}, 167 (2002).

\bibitem{Matthews99} M.R. Matthews, B.P. Anderson,P.C. Haljan, D.S. Hall, C.E. Wieman, and E.A. Cornell, {\it Vortices in a Bose-Einstein condensate}, Phys. Rev. Lett. {\bf 83}, 2498 (1999).

\bibitem{Madison00} K.W. Madison, F. Chevy, W. Wohlleben,  and J. Dalibard, {\it Vortex formation in a stirred Bose-Einstein condensate}, Phys. Rev. Lett. {\bf 84}, 806 (2000).

\bibitem{Weiler08} C.N. Weiler, T.W. Neely, D.R. Scherer,A.S. Bradley, M.J. Davis, and B.P. Anderson, {\it Spontaneous vortices in the formation of Bose-Einstein condensates}, Nature {\bf 455} 948-U37 (2008).

\bibitem{Freilich10} D.V. Freilich, D.M. Biancho, A.M. Kaufman, T.K. Langin, and D.S. Hall, {\it Real-Time Dynamics of Single Vortex Lines and Vortex Dipoles in a Bose-Einstein Condensate}, Science {\bf 329}, 1182 (2010).

    \bibitem{Neely10} T.W. Neely, E.C. Samson, A.S. Bradley,  J.M. Davis, and B.P. Anderson, {\it Observation of Vortex Dipoles in an Oblate Bose-Einstein Condensate}, Phys. Rev. Lett. {\bf 104}, 160401 (2010).

 \bibitem{Smith04} N.L. Smith, W.H. Heathcote, J.M. Krueger, and C.J. Foot, {\it Experimental observation of the tilting mode of an array of vortices in a dilute Bose-Einstein condensate} Phys. Rev. Lett. {\bf 98}, 080406 (2004).


\bibitem{AboShaeer01} J.R. Abo-Shaer, C. Raman, J.M. Vogels, and W. Ketterle, {\it Observation of vortex lattices in Bose-Einstein condensates}, Science, {\bf 292},476 (2001)

\bibitem{Schweikhard04} V. Schweikhard, I. Coddington, P. Engels, S. Tung, and E.A. Cornell, {\it Vortex-lattice dynamics in rotating spinor Bose-Einstein condensates} Phys. Rev. Lett. {\bf 93}, 210403 (2004).

\bibitem{Coddington03} I. Coddington, P. Engels, V. Schweikhard, and E.A. Cornell, EA {\it Observation of Tkachenko oscillations in rapidly rotating Bose-Einstein condensates}, Phys. Rev. Lett. {\bf 91}, 100402 (2003).

\bibitem{Henn09} E.A.L. Henn, J.A. Seman, G. Roati, K.M.F. Magalhaes, and V.S. Bagnato, {\it Emergence of Turbulence in an Oscillating Bose-Einstein Condensate}, Phys. Rev. Lett. {\bf 103}, 045301 (2009).

\bibitem{Neely13} T.W. Neely, A.S. Bradley, E.C. Samson, S.J. Rooney, E.M. Wright, K.J.H. Law, R. Carretero-Gonzalez, P.G. Kevrekidis, M.J. Davis, and B.P. Anderson, {\it Characteristics of Two-Dimensional Quantum Turbulence in a Compressible Superfluid}, Phys. Rev. Lett. {\bf 111}, 235301 (2013).

\bibitem{Lamb} H. Lamb, {\it Hydrodynamics}, (Cambridge University Press 1895)

\bibitem{PopovBook} V. N. Popov, {\it Functional integrals and in quantum field theory and statistical physics} (Kluwer, Boston, 1983).

\bibitem{Hydrodynamics} L.M. Milne-Thomson, {\it Theoretical Hydrodynamics} (Dover, 1996).

\bibitem{Ames1992} W.F. Ames, {\it Numerical methods for partial differential equations}, Academic
Press, San Diego, 3rd. edition, (1992).

\bibitem{Adler1984} S. L. Adler and T. Piran, Rev. Mod. Phys., {\bf 56} 1–40, (1984).

\bibitem{ThoulessAnglin2007} D.~J.~Thouless and J.~R.~Anglin, Vortex mass in a superfluid at low
frequencies. Phys. Rev. Lett., {\bf 99} 105301 (2007).

\bibitem{We1} A. Klein, I. Aleiner, and O. Agam (in preparation).

\bibitem{Feynmann55} R. P. Feynman,  Application of quantum mechanics to liquid helium. In
C. J. Gorter, editor, {\it Progress in Low Temperature Physics}, volume 1.
North-Holland, Amsterdam, (1955).

\bibitem{Popov73} V.~N.~Popov, Quantum vortices and phase transitions in Bose systems. Sov.
Phys. JETP, {\bf 37} 341, (1973).


\bibitem{DuanLeggett92} J.~M.~Duan and A.~J.~Leggett, Inertial mass of a moving singularity in a
Fermi superuid. Phys. Rev. Lett., {\bf 68} 1216 (1992).

\bibitem{Sonin97} E.B. Sonin, {\it Magnus force in superfluids and superconductors}, Phys. Rev. B {\bf 55}, 485 (1997).

\bibitem{Dykman12} M. Dykman  {\it Fluctuating nonlinear oscillators} (Oxford, 2012).

\bibitem{SpinorSuperfluidReview} Y. Kawaguchi and M. Ueda, {\it  Spinor Bose-Einsten condensates}, Phys. Rep.  {\bf 520}, 253 (2012).

\bibitem{Saad2003} Y. Saad, {\it Iterative methods for sparse linear systems}, (Society for Indus-
trial and Applied Mathematics, 2003).


\end{thebibliography}
\end{document}